\documentclass[
 reprint,
 superscriptaddress,
 amsmath,amssymb,
 floatfix,
 prl
]{revtex4-2}

\usepackage[utf8]{inputenc}
\usepackage[T1]{fontenc}
\usepackage{graphicx}
\usepackage{amsmath,amssymb,amsfonts,amsthm,bm,braket}
\usepackage[colorlinks=true,citecolor=blue,urlcolor=blue,linkcolor=blue]{hyperref}

\newcommand{\Tr}{\operatorname{Tr}}
\newcommand{\id}{\operatorname{id}}
\newcommand{\Ens}{\mathsf{Ens}}
\newcommand{\cD}{\mathcal D}
\newcommand{\cR}{\mathcal R}
\newcommand{\cL}{\mathcal L}

\newcommand{\cA}{\mathcal A}

\newcommand{\cW}{\mathcal W}

\newcommand{\Ep}{E_{\rm p}^{H_B}}
\newcommand{\supp}{\operatorname{supp}}

\theoremstyle{definition}
\newtheorem{theorem}{Theorem}
\newtheorem{proposition}{Proposition}

\newtheorem{lemma}{Lemma}

\newcounter{endmattersection}
\renewcommand{\theendmattersection}{\Roman{endmattersection}}

\begin{document}

\title{Remote Use of Quantum Machines in the Adversarial Scenario}

\author{Mir Alimuddin}
\email{aliphy80@gmail.com}
\affiliation{ICFO-Institut de Ciencies Fotoniques, The Barcelona Institute of Science and Technology, Av. Carl Friedrich Gauss 3, 08860 Castelldefels (Barcelona), Spain}
\affiliation{Department of Physical Sciences, Bose Institute, EN 80, Bidhannagar, Kolkata 700091, India}

\author{Leonardo Zambrano}
\affiliation{ICFO-Institut de Ciencies Fotoniques, The Barcelona Institute of Science and Technology, Av. Carl Friedrich Gauss 3, 08860 Castelldefels (Barcelona), Spain}

\author{Antonio Ac\'in}
\affiliation{ICFO-Institut de Ciencies Fotoniques, The Barcelona Institute of Science and Technology, Av. Carl Friedrich Gauss 3, 08860 Castelldefels (Barcelona), Spain}
\affiliation{ICREA -- Instituci\'o Catalana de Recerca i Estudis Avan\c{c}ats, Llu\'is Companys 23, 08010 Barcelona, Spain}

\begin{abstract}
Networked quantum technologies allow users to operate quantum machines
hosted by remote servers, but classical commands alone cannot ensure
that only the intended user can prepare the useful inputs that activate
them. We formulate the problem of adversarial remote use of quantum machines in the framework of
quantum resource theories. A device controlled by Bob is correlated to a remote user, Alice, through a bipartite state, and also to an adversary Eve holding
its purification. Alice and Eve compete to programme Bob's machine by remotely
preparing its inputs: the conditional state carries the quantum
resource, while the measurement outcome selects the machine operation.
Remote use is truly advantageous when Alice is guaranteed to unlock a larger average task value than the adversary Eve, given the same marginal state of Bob. We show that for tasks with a convex figure of merit, 
the competition is governed by
complementary upper roofs and, for spectral tasks, where the figure of merit is invariant under unitary operations, by a corresponding
entanglement imbalance. Purification, work extraction, quantum
batteries, and coherence distillation show that the same noisy state
may secure exclusivity one machine but not another. Finally, we order Alice's and
Eve's complete programming powers through a task-independent simulation
preorder, based on the ensembles each party can prepare on Bob's side.
Shared quantum correlations thereby become a key resource for controlling
access to remote quantum machines.
\end{abstract}

\maketitle

\textit{Introduction.--}
Cloud-based quantum processors~\cite{preskill2018quantum,devitt2016performing,wehner2018quantum}, networked quantum sensors~\cite{komar2014quantum}, and server-side quantum machines point toward a future in which users operate devices they do not physically possess. Such devices may instead be hosted by a remote server and activated at a distance. This raises a fundamental question: can quantum mechanics enable an authorized remote user to operate such a device more
efficiently than an adversary?

We address this question for machines whose performance depends on the
resource content of their input states.  Bob hosts the machine, while a
distant user Alice prepares its inputs by measuring her share of a
correlated state and communicating the outcome.  The use is \textit{remote} because Alice prepares Bob's input through her local action, and \textit{exclusive} when her optimal task score exceeds that of Eve, who holds the purification of $\rho_{AB}$. The
instruction itself is classical, so any separation between Alice and Eve
must originate from their correlations with Bob.  This objective is related
in spirit to quantum key distribution~\cite{gisin2002quantum,scarani2009security},
certified randomness~\cite{pironio2010randomness,acin2016}, and blind
quantum computation~\cite{broadbent2009universal,fitzsimons2017private}, but
asks a different operational question: who can prepare the more useful
inputs for Bob's machine?

Quantum resource theories provide a common language for this
comparison~\cite{coecke2016mathematical,chitambar2019quantum} by specifying free operations and quantifying the operational value of resourceful inputs.  While measurement-induced enhancements have been studied for coherence, work extraction, and ergotropy~\cite{chitambar2016assisted,morris2019assisted,francica2017daemonic}, and an adversarial setting was recently introduced for memory erasure~\cite{alimuddin2025}, we elevate this competition to arbitrary tasks. Physically, measuring a correlated subsystem fine-grains Bob's fixed marginal state into a conditional ensemble, allowing him to tailor his machine's free operations to specific inputs. Exclusivity strictly depends on whether Eve can match this enhanced capability. The relevant objects are therefore the complete sets of
Bob-side ensembles remotely preparable by Alice and Eve.

After formalizing the scenario, we derive results that apply to any convex resource theories, including purification, thermal work, quantum batteries, and coherence. However, beyond the individual tasks, we introduce a simulation preorder for remote usage:
no advantage can be obtained by Alice in any task whenever each of the ensembles she can 
prepare can equally be prepared by Eve. This property of exact ensemble inclusion 
is characterized by a quantum statistical
Morphism \cite{buscemi2012comparison,buscemi2016degradable} and, more restrictively, positive and completely positive trace preserving (PTP and CPTP) maps in the corresponding direction provide sufficient conditions for this inclusion.

{\it Remote quantum machines and adversarial setting.--} We describe Bob's machine within a resource theory $F$ by
\begin{align}
\mathcal M_F
=
\bigl(
\mathcal H_B,\mathcal S_F,\mathcal O_F,\mathcal R
\bigr) \nonumber,
\end{align}
where
$\mathcal H_B$ is its input space,
$\mathcal S_F\subseteq\mathcal D(\mathcal H_B)$ is the set of free
states, $\mathcal O_F\subseteq{\rm CPTP}(\mathcal H_B)$ is the family
of allowed free operations, and $\cR:D(\mathcal H_B)\rightarrow \mathbb{R}$ 
provides the optimal task score
attainable from a given input state. Free operations cannot generate the
resource from a free state:
$\Lambda_F(\rho_F)\in\mathcal S_F$ for every
$\rho_F\in\mathcal S_F$ and $\Lambda_F\in\mathcal O_F$, although they
may consume, preserve, or transform a resource already present in the
input. Since $\cR$ quantifies the optimal performance of the machine,
$\cR(\Lambda_F(\omega_B))\leq\cR(\omega_B)$.
Throughout, we work in the convex setting: probabilistic mixtures of
free states and free operations remain free, and
$\cR(\sum_i p_i\omega_i)\leq\sum_i p_i\cR(\omega_i)$. Physically,
forgetting the classical label $i$ replaces the ensemble
$\{p_i,\omega_i\}_i$ by its average state and cannot create a resource
or improve the machine's average performance. The machine itself
therefore supplies no resource; its functionality is determined by the
resource carried by its input and by how that input is processed using
$\mathcal O_F$.

Remote use exploits precisely the distinction between a labeled
ensemble and its average state. If Alice holds a purification of
$\rho_B$, the Hughston-Jozsa-Wootters (HJW) theorem~\cite{hughston1993complete} allows her to
prepare any ensemble decomposition of $\rho_B$ by measuring her system
and communicating the outcome to Bob. Because the
shared state is pure, no third party possesses correlated side
information about Bob's input. For a mixed shared state $\rho_{AB}$,
however, Alice generally loses access to some ensemble decompositions,
while the purifying system acquires nontrivial control over Bob's
conditional states. We formalize this by assigning the purification to untrusted adversarial Eve:
\begin{align}
    \rho_{AB}
 &=\Tr_E|\Psi\rangle\!\langle\Psi|_{ABE}, &
    \rho_{EB}
    &=\Tr_A|\Psi\rangle\!\langle\Psi|_{ABE}.
\end{align}

\begin{figure}[t]
    \centering
    \includegraphics[width=1.0\columnwidth]{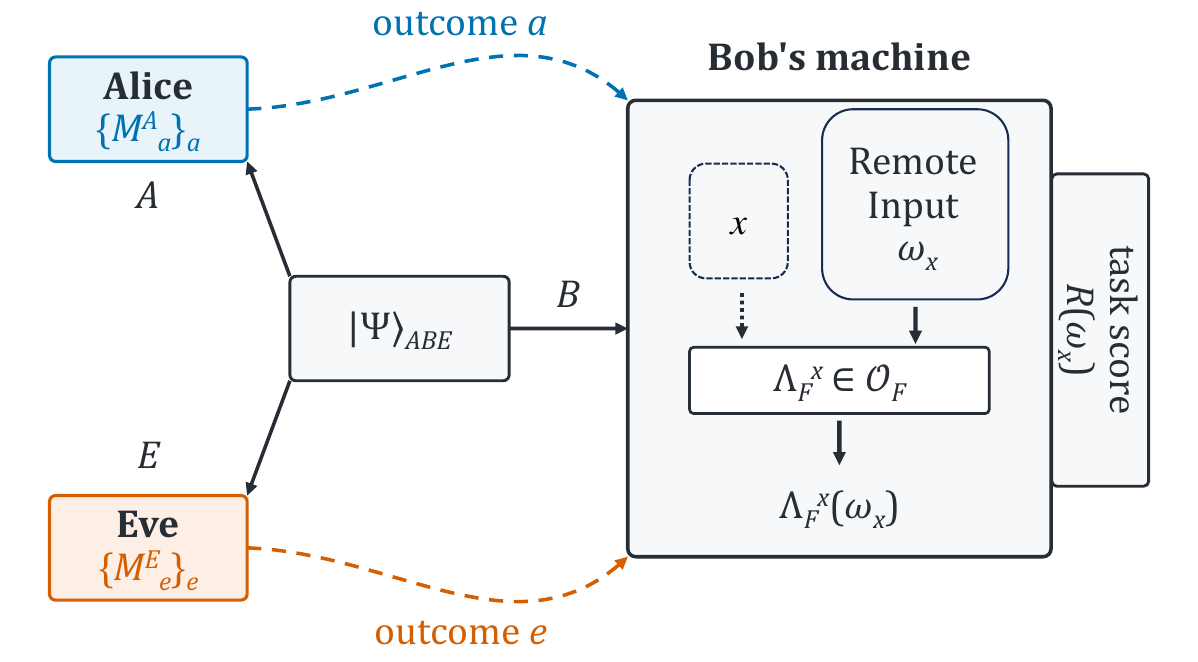}
\caption{\textbf{Exclusive remote usability.}
Alice and Bob share $\rho_{AB}$, while Eve holds the purifying system
$E$ of $|\Psi\rangle_{ABE}$. In the two competing strategies, Alice or
Eve measures her subsystem and communicates the outcome $x=a$ or
$x=e$, thereby remotely preparing the conditional state $\omega_x$ on
Bob's system. Each strategy supplies a two-part program for Bob's
machine: $\omega_x$ is the quantum input on which its performance
depends, while $x$ is the classical instruction selecting a free
operation $\Lambda_F^x\in\mathcal O_F$. The operation produces the
output state $\Lambda_F^x(\omega_x)$, and optimization over the allowed
operations gives the task score $\cR(\omega_x)$. The classical
instruction carries no quantum resource by itself, but makes the
fine-grained quantum input operationally useful. Both remotely prepared
ensembles average to the same marginal $\rho_B$. Alice has a exclusive
advantage when her optimal average task score strictly exceeds Eve's.}
    \label{fig:secure-remote-machine}
\end{figure}

Alice and Eve can now both remotely prepare ensembles on $B$ by
measuring their respective systems. For $X\in\{A,E\}$, define
\begin{align}
\label{eq:remote-ensembles}
\Ens_{X\to B}(\rho_{XB})
= \Big\{&\{p_x,\omega_x\}_x : \exists\{M_x^X\}_x\ \text{such that} \nonumber\\
&p_x=\Tr[(M_x^X\otimes\mathbb I_B)\rho_{XB}]>0, \\
&\omega_x= \Tr_X[(M_x^X\otimes\mathbb I_B)\rho_{XB}]/p_x \Big\}\nonumber\,,
\end{align}
where $\{M_x^X\}_x$ denotes a measurement on $X$.
Every such ensemble averages to $\rho_B$. As shown in Fig.~\ref{fig:secure-remote-machine}, each strategy programs Bob's machine through a joint quantum-classical instruction. The conditional state $\omega_x$ supplies the quantum resource bounding the machine's performance, while the classical outcome $x$ selects the optimal free operation $\Lambda_F^x\in\mathcal O_F$. The classical message therefore carries no resource itself, but is strictly necessary to make the fine-grained quantum input operationally useful.

These measurement-induced ensemble sets are closely related to
steering assemblages~\cite{wiseman2007steering} and quantify Alice's
and Eve's competing abilities to program Bob's machine. Since the
isometry relating any two purifications of $\rho_{AB}$ can be absorbed
into Eve's measurement, $\Ens_{E\to B}(\rho_{EB})$ is independent of
the chosen purification. Exclusive remote usability therefore asks
whether Alice can program the machine more effectively than an
adversary holding all side information compatible with
$\rho_{AB}$. We now study the conditions under which a state $\rho_{AB}$ provides a strict exclusive advantage for different convex task scores $\mathcal{R}$.

{\it Task-dependent exclusive advantage.--}
Remote preparation is useful only when the classical outcome allows Bob
to outperform what he can achieve from the unlabeled average state
$\rho_B$. For either party $X\in\{A,E\}$, we quantify this improvement
by the remote assisted gain
\begin{align}
\label{assistance}
    \cA_{X\to B}^{\cR}(\rho_{XB})
    \! = \!\!\!\!\!
    \sup_{\{p_x,\omega_x\}\in\Ens_{X\to B}(\rho_{XB})}
    \sum_x p_x\cR(\omega_x)
    -
    \cR(\rho_B).
\end{align}
Classical correlations may suffice to have a non-zero gain.

Because $\cR$ is convex, resolving a measurement outcome into finer
outcomes cannot decrease the average score. The optimization may
therefore be restricted to rank-one measurements. By the HJW
theorem~\cite{hughston1993complete,bennett1996mixed,
divincenzo1999assistance}, such measurements by Alice generate exactly
the pure-state decompositions of $\rho_{EB}$, whereas those by Eve
generate the pure-state decompositions of $\rho_{AB}$. Accordingly,
for a bipartite state $\sigma_{XB}$ and score $\cR$, define the upper roof
\begin{align}
\label{eq:upper_roof}
    \widehat{\cR}(\sigma_{XB})
    =
    \sup_{\{p_k,\psi_k\}}
    \sum_k p_k\,
    \cR\!\left(
        \Tr_X|\psi_k\rangle\!\langle\psi_k|
    \right),
\end{align}
where the supremum is over all pure-state decompositions of
$\sigma_{XB}$. The assisted gains then take the complementary form
\begin{align}
\label{eq:complementary_assistance}
    \cA_{A\to B}^{\cR}(\rho_{AB})
    &=
    \widehat{\cR}(\rho_{EB})-\cR(\rho_B),
    \nonumber\\
    \cA_{E\to B}^{\cR}(\rho_{EB})
    &=
    \widehat{\cR}(\rho_{AB})-\cR(\rho_B).
\end{align}

{\bf Definition (exclusive advantage).--}
For a task score $\cR$, define the exclusive advantage of Alice over Eve as
\begin{align}
\label{secure-advantage}
    \Delta^{\cR}(\rho_{AB})
    &:=
    \cA_{A\to B}^{\cR}(\rho_{AB})
    -
    \cA_{E\to B}^{\cR}(\rho_{EB})
    \nonumber\\
    &=
    \widehat{\cR}(\rho_{EB})
    -
    \widehat{\cR}(\rho_{AB}).
\end{align}
We say that $\rho_{AB}$ provides Alice with a exclusive advantage
for $\cR$ whenever $\Delta^{\cR}(\rho_{AB})>0$. The exclusive advantage is therefore determined entirely by which party
can unlock the larger average task score from Bob's fixed marginal
state. While ordinary remote assistance does not require
entanglement, having a exclusive advantage is more
demanding, as it will become clear below. For every pure entangled state, Eve is uncorrelated with Bob, guaranteeing Alice exclusivity for any strictly convex task score; we show that this advantage persists under sufficiently weak noise (see Appendix \ref{app:robustness} for details).

\begin{theorem}
\label{thm:robustness}
Let $|\Phi\rangle_{AB}$ be pure and entangled, and let $\cR$ be a
continuous, strictly convex functional. Then, there exists $\epsilon^*>0$ such that
\begin{align}
    \Delta^{\cR}(\rho_{AB})>0
    \quad\text{whenever}\quad
    \|\rho_{AB}-\Phi_{AB}\|_1\leq\epsilon^*.
\end{align}
\end{theorem}

Thus far, $\cR$ has been assumed only to be convex. For a generic
convex task score, the upper roofs (Eq. \ref{eq:upper_roof}) have a direct
operational meaning but need not be entanglement measures. Such an
interpretation emerges when $\cR$ is also spectral, or, equivalently,
unitarily invariant. In that case, $\cR$ takes the same value $\cR_{\rm pure}$ on every pure state of $B$. The quantity $e_{\cR}(\vert{}\psi\rangle_{XB}) = \cR_{\rm pure}-\cR(\psi_B)$ is therefore nonnegative and Schur-concave in the Schmidt spectrum, defining a pure-state entanglement monotone~\cite{vidal2000monotones}. 
Its convex-roof extension satisfies
\begin{align}
    E_{\cR}(\sigma_{XB})
    & =
    \inf_{\{p_k,\psi_k\}}
    \sum_k p_k\,
    e_{\cR}(|\psi_k\rangle_{XB}) \\
    & =\cR_{\rm pure}-\widehat{\cR}(\sigma_{XB}).
\end{align}
Substitution into Eq.~\eqref{secure-advantage} gives the following
criterion.
\begin{theorem}
\label{thm:spectral-roof}
Let $|\Psi\rangle_{ABE}$ purify $\rho_{AB}$, with
$\rho_{EB}=\Tr_A|\Psi\rangle\!\langle\Psi|_{ABE}$. For every convex
spectral $\cR$,
\begin{align}
    \Delta^{\cR}(\rho_{AB})>0
    \quad\Longleftrightarrow\quad
    E_{\cR}(\rho_{AB})>E_{\cR}(\rho_{EB}).
\end{align}
\end{theorem}
Thus, for a spectral task, Alice has a strict exclusive advantage exactly
when she shares more task-induced entanglement with Bob than Eve does;
see Appendix~D.

{\it Applications: remote quantum machines.--}
We now ask which specific machines Alice can operate with an exclusive advantage. Consider a purifier, a thermal engine, a quantum battery, and a coherence distiller. While all four machines receive the exact same marginal state $\rho_B$, they are programmed through different conditional ensembles, revealing precisely which physical feature of the shared correlations supplies the exclusive advantage.

\textit{a) Exclusive purifier.--}
Consider first a machine that distills pure states from Bob's input.
In the resource theory of purity, the task score
\begin{align}
\mathcal P(\omega_B)=\log d_B-S(\omega_B)
\end{align}
gives the asymptotic distillable purity under noisy
operations~\cite{horodecki2003reversible}. By resolving the mixed state
$\rho_B$ into purer conditional inputs, Alice can increase the
purifier's average yield. Her advantage over Eve is
\begin{align}
\Delta^{\mathcal P}(\rho_{AB})
=
E_F(\rho_{AB})-E_F(\rho_{EB}),
\end{align}
where $E_F$ is the entanglement of formation
\cite{bennett1996mixed}. Thus the resource that guarantees exclusivity to the remote
purifier is precisely the imbalance of entanglement of formation:
Alice operates it exclusively if she shares more of this entanglement with
Bob than Eve does.

\textit{b) Exclusive thermal engine.--}
Suppose Bob's machine extracts work from a system with
Hamiltonian $H_B$ in contact with a heat bath at temperature $T$. Its
asymptotic work value is given by the difference in nonequilibrium free energy between system and the corresponding thermal state:
\begin{align}
    \cW_\beta(\rho_B)
    &=
    k_BT\,D(\rho_B\Vert\gamma_B^\beta)
    \nonumber\\
    &=
    \Tr(H_B\rho_B)-k_BT S(\rho_B)+k_BT\ln Z_B
\end{align}
with $\beta=(k_BT)^{-1}$
\cite{brandao2013resource,horodecki2013fundamental,
morris2019assisted}. Alice's classical instruction allows Bob to adapt
the extraction protocol to the remotely prepared conditional state.

Although the work value depends on both energy and entropy, every
remotely prepared ensemble has average state $\rho_B$ and therefore the
same average energy. The energy contribution cannot distinguish Alice
from Eve and cancels from their comparison. Exclusivity is consequently determined entirely by their respective abilities to reduce the conditional entropy, yielding the following advantage in remotely extractable work:
\begin{align}
\Delta^{\mathcal W_\beta}(\rho_{AB})
=
k_BT
\bigl[
E_F(\rho_{AB})-E_F(\rho_{EB})
\bigr].
\end{align}
The same entanglement imbalance that guarantees exclusivity in purification also guarantees exclusivity in bath-assisted work extraction, despite the two machines performing physically distinct tasks.

\begin{figure}[t]
    \centering
\includegraphics[width=0.95\columnwidth]{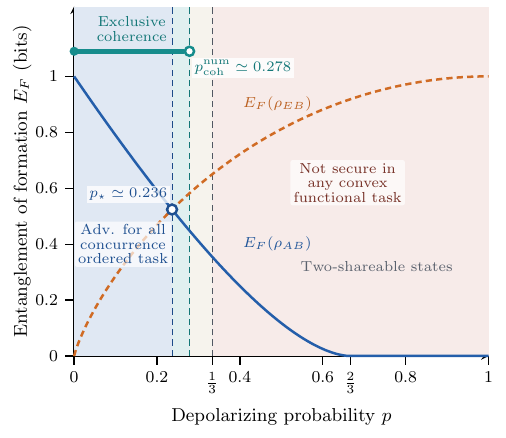}
    \caption{\textbf{Task-dependent usefulness of two-qubit Werner states.}
Here $p$ is the depolarizing-noise strength, and the solid and dashed
curves show $E_F(\rho_{AB})$ and $E_F(\rho_{EB})$, respectively.
Alice has an exclusive advantage for every concurrence-ordered
spectral task when $p<p_\star\simeq0.23629$, while her coherence
advantage persists up to
$p_{\rm coh}^{\rm num}\simeq0.27824$; see End Matter
Sec.~\ref{sec:end_task_dependent} and the Appendix. For $p\geq1/3$,
two-shareability allows Eve to simulate Alice, ruling out exclusivity
for any task. Whether another task exhibits exclusivity for
$p_{\rm coh}^{\rm num}\leq p<1/3$ remains open.}
    \label{fig2}
\end{figure}

\textit{c) Exclusive quantum battery.--}
The situation changes when Bob uses the input as a quantum battery and
extracts work through cyclic unitary control, without access to a heat
bath. The relevant task score is given by the ergotropy~\cite{allahverdyan2004maximal}
\begin{align}
\mathcal W_{\rm erg}(\rho_B)
=
\Tr(H_B\rho_B)-E_{\rm p}^{H_B}(\rho_B),
\end{align}
where $E_{\rm p}^{H_B}(\rho_B)$ is the energy of the passive state
associated with $\rho_B$. Once again, the average energy is fixed by
$\rho_B$ and cancels between Alice and Eve. What distinguishes their
performances now is how effectively each party can remotely program Bob's system into a lower energetic passive state. 
Let the convex roof of the passive
energy be defined as
\begin{align}
    \mathcal E_{\rm p}(\rho_{XB})
    =
    \inf_{\{p_k,\psi_k\}}
    \sum_k p_k
    \Ep\!\left(
        \Tr_X|\psi_k\rangle\!\langle\psi_k|
    \right).
\end{align}
The exclusive ergotropic advantage is
\begin{align}
\Delta^{\mathcal W_{\rm erg}}(\rho_{AB})
=
\mathcal E_{\rm p}(\rho_{AB})
-
\mathcal E_{\rm p}(\rho_{EB}).
\end{align}
The security resource is therefore still an entanglement roof, but one
that depends explicitly on the battery Hamiltonian and is inequivalent to $E_F$ in higher dimensions \cite{Alimuddin2020,yang2024ergotropicgap}. Consequently, a shared state
may fail to produce exclusive bath-assisted work extraction while still enabling
Alice to charge or discharge Bob's quantum battery exclusively. An
explicit example is given in the End Matter Sec.~\ref{sec:end_task_dependent}.

\textit{d) Exclusive coherence distiller.--}
Finally, consider a machine that distills coherence relative to a
fixed incoherent basis of $B$. Under incoherent operations, the
relative entropy of coherence (convex but not spectral)
\begin{align}
C_{\rm r}(\omega_B)
=
S(\Delta_B(\omega_B))-S(\omega_B)
\end{align}
gives the asymptotic rate of distilling maximally coherent
states~\cite{baumgratz2014quantifying,winter2016operational}. Here
$\Delta_B$ denotes complete dephasing in the chosen basis. Exclusivity is not determined anymore only by the spectral properties of the state, as
the term $S(\Delta_B(\omega_B))$ depends on the eigenvalues of
$\omega_B$, but also on the orientation of the state relative to the
incoherent basis. This basis-dependent contribution does not cancel between
Alice and Eve. Their exclusive advantage in terms of upper roof as Eq.~\eqref{eq:upper_roof} is given by
\begin{align}
\Delta^{C_{\rm r}}(\rho_{AB})
=
\widehat C_{\rm r}(\rho_{EB})
-
\widehat C_{\rm r}(\rho_{AB}),
\end{align}
where both entropy contributions in each $\widehat C_{\rm r}$ must be optimized jointly over the
same decomposition. 

Figure~\ref{fig2} illustrates this task dependence for the same noisy
Werner family: machines governed by spectral scores and the
nonspectral coherence distiller have different exclusive regions. Since
locating each boundary requires a separate complementary-roof
optimization, we now ask whether usefulness or uselessness can instead
be decided independently of the task by comparing Alice's and Eve's
full ensemble-preparation capability.

{\it Task-independent simulation preorder.--}
Rather than optimizing a particular task score, we now compare the full
menus of quantum programs available to Alice and Eve. Each remotely
prepared ensemble specifies both the conditional inputs to Bob's machine
and the classical labels selecting its operations. If
$\Ens_{E\to B}(\rho_{EB})\subseteq\Ens_{A\to B}(\rho_{AB})$, Alice can
reproduce every program available to Eve and is therefore never worse
for any machine. The reverse inclusion rules out a strict exclusive
advantage for Alice in every task. If neither inclusion holds, their
programming powers are incomparable and the outcome depends on the
machine being operated.

A familiar sufficient mechanism for these inclusions is a physical
degrading channel. For example, if a CPTP map $\cD_{E\to A}$ satisfies $\rho_{AB} =(\cD_{E\to A}\otimes\id_B)(\rho_{EB}) $
Eve can simulate any Alice's measurement by pulling it back through
$\cD_{E\to A}^{\dagger}$, where $\dagger$ denotes the adjoint map. CPTP degradability therefore implies that
Eve can match Alice for every machine and the state $\rho_{AB}$ provides no advantage for any task.
Separable states are examples of those states, but also entangled states $\rho_{AB}$ that are two-shareable, that is, 
for which there exists $\rho_{ABE}$ such that $\Tr_A\rho_{ABE}=\rho_{AB}$. Crucially, physical degradability 
is not the exact condition, but rather a comparison of quantum
statistical models: every measurement strategy of Alice (Eve) must be
simulable by Eve (Alice) while reproducing, outcome by outcome,  the same
conditional states on Bob's system~\cite{blackwell1953equivalent,shmaya2005comparison,
buscemi2012comparison,buscemi2016degradable}.

\begin{theorem}
\label{thm:bob-ensemble-simulation}
Let $|\Psi\rangle_{ABE}$ purify $\rho_{AB}$, with
$\rho_{EB}=\Tr_A|\Psi\rangle\!\langle\Psi|_{ABE}$. Then
$\Ens_{E\to B}(\rho_{EB})\subseteq\Ens_{A\to B}(\rho_{AB})$ if and
only if there exists a Hermiticity- and trace-preserving statistical
morphism $\cL_{A\to E}$ for the model induced by $\rho_{AB}$ such that
\begin{align}
    \rho_{EB}
    =
    (\cL_{A\to E}\otimes\id_B)(\rho_{AB}).
\end{align}
\end{theorem}

The proof is given in Appendix~G. The theorem therefore extends universal simulation beyond physical
degradability: even when no CPTP map from $A$ to $E$ exists, a
statistical morphism can allow Alice to reproduce every program
available to Eve and hence never perform worse in any task. Such a
morphism needs to act correctly only on the operator model probed by
$\rho_{AB}$ and needs not to be positive on arbitrary inputs. A PTP map is
a stronger realization, whose adjoint maps every measurement on $E$ to a valid measurement on $A$,
while a CPTP map is even stronger and represents a physical degrading
channel. Thus,

{\footnotesize
\begin{align}
    \text{CPTP deg.}
    \ \Longrightarrow\
    \text{PTP deg.}
    \ \Longrightarrow\
    \Ens_{E\to B}\subseteq\Ens_{A\to B}.
\end{align}
}
with the last condition characterized exactly by the theorem. 
None of the reverse implications is true. In fact, 
the End Matter Sec.~\ref{sec:end_task_independent} presents an entangled example in which Alice simulates Eve
despite the absence of a CPTP degrading map, and evaluates the
resulting exclusive advantages for the machines considered above.

For rank-two two-qubit states, this gap closes: the
degrading--antidegrading dichotomy makes the CPTP criterion exact. The
amplitude-damping family illustrates the resulting order directly:
Alice simulates Eve below the self-complementary point, Eve simulates
Alice above it, and their programming powers coincide at this point;
see End Matter Sec.~\ref{sec:end_task_independent} and Appendix for details.

\textit{Discussion.--}
A remotely hosted quantum machine receives two forms of programming:
a quantum input supplied by Bob's conditional state and a classical
instruction selecting how that input is processed. Classical
communication alone cannot make this use exclusive. The advantage arises
from the correlations that determine which conditional quantum inputs
Alice and Eve can prepare.

Our results separate two levels of this exclusivity. At the
machine-dependent level, entanglement is always necessary, but the
quantity that controls performance depends on the physical task.
Exclusivity of purification and thermal work are guranteed by the entanglement of
formation; battery work is controlled by a Hamiltonian-dependent
entanglement roof; and coherence distillation additionally probes the
basis-dependent geometry of the remotely prepared states. The same
shared state can therefore give Alice an exclusive advantage for one quantum machine while providing none for another.

The task-independent preorder removes the need to choose a machine:
ensemble inclusion means that one party can reproduce the other's
entire menu of quantum programs and hence match its performance for
every task. Its exact characterization by a statistical morphism
separates operational simulation from physical state conversion,
whereas PTP and CPTP degradability provide successively stronger
realizations. This hierarchy offers a unified way to classify shared
correlations as universally useless, universally ordered, or
intrinsically task dependent, and establishes quantum correlations as
a resource for controlling access to remotely programmed quantum
devices.

\textit{Acknowledgments.--} M.A., L.Z., and A.A.\ were supported by 
the Government of Spain (Severo Ochoa CEX2019-000910-S, FUNQIP, and European Union NextGenerationEU PRTR-C17.I1), 
Fundaci\'o Cellex, 
Fundaci\'o Mir-Puig,
Generalitat de Catalunya (CERCA program), 
the European Union (QSNP, 101114043; QuantERA project Veriqtas), 
the ERC Advanced Grant CERQUTE, 
and the AXA Chair in Quantum Information Science.
M.A.\ additionally acknowledges funding from the European Union (QURES, 101153001).

\vspace{0.3cm}
\clearpage

\vspace{0.3cm}
\clearpage

{\Large\textbf{End Matter}}

\vspace{0.2cm}
\refstepcounter{endmattersection}
\label{sec:end_task_dependent}
\noindent
{\it \theendmattersection. Task-dependent structure.--}
For a pure two-qubit state, the entanglement functional induced by any
spectral task score is a function $f_{\cR}(C)$ of the concurrence $C$, or equivalently 
the smallest Schmidt coefficient.
Whenever $f_{\cR}$ is convex and strictly increasing, Wootters'
equal-concurrence construction gives, for every two-qubit state
$\sigma$, not necessarily pure, 
\begin{align}
    E_{\cR}(\sigma)
    =
    f_{\cR}\!\left(C(\sigma)\right).
    \label{eq:end_flat_roof}
\end{align}
Such roofs are therefore strictly increasing functions of the
entanglement of formation
\begin{align}
    E_F(\sigma)
    =
    h_2\!\left(
        \frac{1+\sqrt{1-C(\sigma)^2}}{2}
    \right)
\end{align}
and induce the same ordering~\cite{wootters1998}. The convexity of
$f_{\cR}$ is essential; the proof and its dimensional qualifications
are given in Appendix~\ref{app:two_qubit_reduction}.

The noisy family in Fig.~\ref{fig2} possesses additional symmetry that
makes the spectral ordering analytically tractable even though its
complementary state is not two-qubit. Consider
\begin{align}
    \rho_{AB}(p)
    =
    (1-p)|\Phi^+\rangle\!\langle\Phi^+|
    +
    p\frac{\mathbb I_{AB}}{4},
    \qquad 0\leq p\leq1.
    \label{eq:end_werner}
\end{align}
Its concurrence is
\begin{align}
    C_{AB}
    =
    \max\left\{0,1-\frac{3p}{2}\right\},
\end{align}
and Wootters' formula gives
\begin{align}
    E_F(\rho_{AB})
    =
    h_2\!\left(
        \frac{1+\sqrt{1-C_{AB}^{\,2}}}{2}
    \right).
    \label{eq:end_werner_eof_ab}
\end{align}

A minimal purification has $\dim E=4$, so Wootters' two-qubit formula
cannot be applied directly to $\rho_{EB}$. Instead, its entanglement
roof is fixed by the isotropy of $\rho_{AB}(p)$. A rank-one measurement
on $A$ with Bloch direction $\boldsymbol n$ prepares the normalized
state
\begin{align}
    \omega_{\boldsymbol n}
    =
    \frac12
    \left[
        \mathbb I_B
        +(1-p)\widetilde{\boldsymbol n}\cdot\boldsymbol\sigma
    \right]
    \label{eq:end_werner_conditional_state}
\end{align}
on $B$, where
$|\widetilde{\boldsymbol n}|=1$. Its spectrum is therefore always
\begin{align}
    \operatorname{spec}(\omega_{\boldsymbol n})
    =
    \left(1-\frac p2,\frac p2\right),
    \label{eq:end_werner_fixed_spectrum}
\end{align}
independently of the measurement direction. By the HJW theorem, every
pure-state decomposition of $\rho_{EB}$ is generated by such a
rank-one POVM on $A$. Hence every pure component has the same Bob-side
Schmidt spectrum, and the complementary spectral roof is flat. In
particular,
\begin{align}
    E_F(\rho_{EB})
    =
    h_2\!\left(\frac p2\right).
    \label{eq:end_werner_eof_eb}
\end{align}

More generally, let $f_{\cR}(C)$ be the pure-state entanglement
functional induced by a spectral task score. The fixed spectrum above
corresponds to
\begin{align}
    C_E=\sqrt{2p-p^2},
\end{align}
and therefore
\begin{align}
    E_{\cR}(\rho_{EB})
    =
    f_{\cR}(C_E).
\end{align}
If $f_{\cR}$ is convex and strictly increasing, Wootters'
equal-concurrence construction also gives
$E_{\cR}(\rho_{AB})=f_{\cR}(C_{AB})$. Consequently,
\begin{align}
    \Delta^{\cR}>0
    \quad\Longleftrightarrow\quad
    C_{AB}>C_E
    \quad\Longleftrightarrow\quad
    p<p_\star,
\end{align}
where
\begin{align}
    p_\star
    =
    \frac{10-4\sqrt3}{13}
    \simeq0.236292.
\end{align}
Thus, for this family, all spectral tasks whose induced pure-state
entanglement measures are convex and strictly increasing functions of
concurrence share the same exclusive-advantage region. This coincidence follows from
Wootters' construction on $A{:}B$ together with the isotropic
flat-roof structure on $E{:}B$; it is not a generic property of
full-rank two-qubit states.

Relative-entropy coherence is nonspectral and retains the direction of
$\widetilde{\boldsymbol n}$ relative to the incoherent basis. Its
task-specific roof instead gives exclusive advantage for
$0\leq p<p_{\rm coh}^{\rm num}\simeq0.27824$. At $p=1/3$, the
associated depolarizing channel becomes CPTP antidegradable, allowing Eve
to simulate Alice for all $p\geq1/3$, although
$\rho_{AB}(p)$ remains entangled until $p=2/3$. The interval
$p_{\rm coh}^{\rm num}\leq p<1/3$ may still support security for tasks
not considered here. Analytical details are given in
Appendices~\ref{app:bell_diagonal}.

For the two-qubit Werner family, bath-assisted work extraction and
quantum-battery ergotropy have the same exclusive-advantage region, but
this common ordering does not persist in general scenario or higher dimensions.
Appendix~\ref{app:task_reversal} constructs a flagged tripartite state
for which Eve outperforms Alice in purification and bath-assisted work,
whereas Alice outperforms Eve in battery ergotropy. Thus, the same
correlation can favor different remote users for different machines,
proving that their remotely preparable ensemble sets are incomparable.

\medskip
\refstepcounter{endmattersection}
\label{sec:end_task_independent}
\noindent
{\it \theendmattersection. Task-independent order.--}
CPTP degradability is a sufficient mechanism for
ensemble simulation,
but its failure does not determine the
operational order. Appendix~\ref{app:ptp_not_cptp} constructs
complementary marginals of a pure state on
$A\simeq\mathbb C^8$ and $B\simeq E\simeq\mathbb C^4$ for which a PTP
map $\mathcal Q_{A\to E}$ satisfies
\begin{align}
    \rho_{EB}
    &=
    (\mathcal Q_{A\to E}\otimes\id_B)(\rho_{AB}),
    \nonumber\\[-1mm]
    \Ens_{E\to B}(\rho_{EB})
    &\subsetneq
    \Ens_{A\to B}(\rho_{AB}),
    \label{eq:end_ptp_order}
\end{align}
although no CPTP map implements either conversion direction. The
inclusion is strict, as witnessed by
\begin{align}
    \Delta^{\mathcal P}(\rho_{AB})
    \geq
    \delta
    :=
    0.015648473\ldots>0.
    \label{eq:end_ptp_gap}
\end{align}
Alice is consequently never worse for any machine and strictly wins
purification. If $\mathcal P$ is measured in bits, the corresponding
bath-assisted work satisfies
$\Delta^{\mathcal W_\beta}\geq k_BT\ln2\,\delta$.
This example therefore separates ensemble simulation
from physical state conversion by CPTP. For unrestricted statistical models,
an even larger separation---statistical simulation without any PTP
map---is known~\cite{matsumoto2014example}; whether it can occur
between complementary marginals remains open.

Beyond special families, deciding the simulation hierarchy becomes
substantially more difficult. CPTP convertibility is a semidefinite
feasibility problem: one seeks a positive semidefinite Choi operator
obeying linear trace-preservation and output constraints. PTP
convertibility instead requires the Choi operator to be block positive,
i.e., positive on all product vectors; deciding block positivity is
dual to the separability problem and is NP-hard in
general~\cite{gurvits2003,gharibian2009}. Testing for a statistical
morphism is model dependent, since the map need only act consistently
on the operator system spanned by the state's model~\cite{buscemi2012comparison};
the resulting positivity condition is relative to that operator system,
and no general semidefinite formulation is known.  A useful exception is provided by
rank-two two-qubit states.

\begin{lemma}[Rank-two dichotomy]
\label{lem:end_rank_two_dichotomy}
Let $\rho_{AB}$ be a rank-two two-qubit state and let $E$ be the purification. Then at least one of the following holds:
\begin{align}
    \rho_{EB}
    &=
    (\cD_{A\to E}\otimes\id_B)(\rho_{AB}),
    \label{eq:end_rank_two_degradable}\\
    \rho_{AB}
    &=
    (\cD_{E\to A}\otimes\id_B)(\rho_{EB}),
    \label{eq:end_rank_two_antidegradable}
\end{align}
where the map in the corresponding relation is CPTP.
\end{lemma}
\begin{proof}
Let $|\phi\rangle_{A'B}$ be a purification of $\rho_B$ with
$A'\simeq\mathbb C^2$. Since $|\Psi\rangle_{ABE}$ also purifies
$\rho_B$, there is an isometry $V:A'\to A\otimes E$ with
$|\Psi\rangle_{ABE}=(V\otimes\mathbb I_B)|\phi\rangle_{A'B}$. Hence
\begin{align}
    \rho_{AB}=(\mathcal N\otimes\id_B)(\phi_{A'B}),
    \qquad
    \rho_{EB}=(\mathcal N^c\otimes\id_B)(\phi_{A'B}),
\end{align}
where $\mathcal N=\Tr_E V(\cdot)V^\dagger$ and
$\mathcal N^c=\Tr_A V(\cdot)V^\dagger$ are complementary qubit channels
with a qubit environment, i.e., with at most two Kraus operators. Every
such channel is degradable or
antidegradable~\cite{wolf2007quantum,cubitt2008}; if it has a
single Kraus operator, it is unitary and trivially degradable. Applying
the corresponding relation $\cD_{A\to E}\circ\mathcal N=\mathcal N^c$
or $\cD_{E\to A}\circ\mathcal N^c=\mathcal N$ to $\phi_{A'B}$ yields
Eq.~\eqref{eq:end_rank_two_degradable} or
Eq.~\eqref{eq:end_rank_two_antidegradable}, respectively.
\end{proof}

For the amplitude-damping family, let
$|\chi_q\rangle=|00\rangle+\sqrt q\,|11\rangle$. The complementary
marginals are
\begin{align}
    \rho_{AB}
    &=
    \frac12|\chi_{1-\gamma}\rangle
    \!\langle\chi_{1-\gamma}|
    +
    \frac{\gamma}{2}|01\rangle\!\langle01|,
    \nonumber\\
    \rho_{EB}
    &=
    \frac12|\chi_{\gamma}\rangle
    \!\langle\chi_{\gamma}|
    +
    \frac{1-\gamma}{2}|01\rangle\!\langle01|,
    \label{eq:end_amplitude_states}
\end{align}
with
$C(\rho_{AB})=\sqrt{1-\gamma}$ and
$C(\rho_{EB})=\sqrt{\gamma}$. The corresponding channel compositions
give
\begin{align}
    \Ens_{E\to B}
    &\subseteq
    \Ens_{A\to B},
    &&0\leq\gamma\leq\frac12,
    \nonumber\\[-1mm]
    \Ens_{A\to B}
    &\subseteq
    \Ens_{E\to B},
    &&\frac12\leq\gamma\leq1.
    \label{eq:end_amplitude_order}
\end{align}
At $\gamma=1/2$ the two channels are self-complementary and their
programming powers coincide, as illustrated in
Fig.~\ref{fig:amplitude_damping_order}. Analytical details are given
in Appendix~\ref{app:amplitude_damping}.

\begin{figure}[t]
    \centering
    \includegraphics[
        width=0.98\columnwidth
    ]{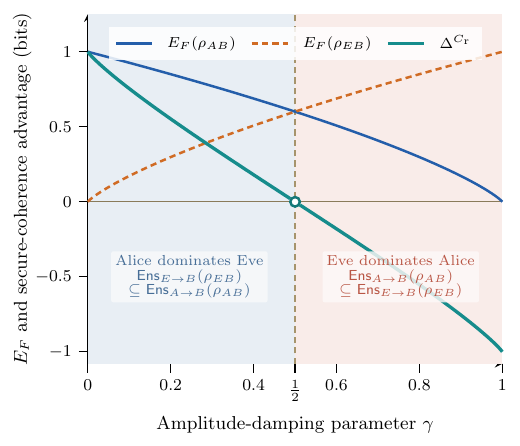}
    \caption{\textbf{Exact simulation preorder for amplitude-damping
states.}
For $0\leq\gamma\leq1$, the solid and dashed curves show
$E_F(\rho_{AB})$ and $E_F(\rho_{EB})$, respectively, whose ordering
determines the signs of the bath-assisted-work and battery-ergotropy
advantages. The teal curve shows the signed coherence advantage
$\Delta^{C_{\rm r}}$. All three advantages change sign at the
self-complementary point $\gamma=1/2$. For this rank-two two-qubit
family, CPTP degradability exactly characterizes the ensemble
preorder. For $\gamma<1/2$,
$\Ens_{E\to B}(\rho_{EB})\subseteq\Ens_{A\to B}(\rho_{AB})$, so Alice
can reproduce every Eve program and is never worse for any machine,
with strict advantages for the tasks shown. For $\gamma>1/2$, the
reverse inclusion holds, ruling out Alice's exclusivity for every
machine. At $\gamma=1/2$, their programming powers coincide.}
    \label{fig:amplitude_damping_order}
\end{figure}


\clearpage
\onecolumngrid

\begin{center}
{\large\bfseries Supplemental Material}
\end{center}

This Supplemental Material supplies the proofs and calculations used in
the main text. All Hilbert spaces are finite dimensional. Unless stated
otherwise, logarithms and von Neumann entropies use the natural base;
$h_2$ and $H_2$ denote binary and Shannon entropies in bits.

\section{A. Complementary-roof representation}
\label{app:complementary_roofs}

We show that, for a convex function $\cR$, optimizing Alice's
measurement is equivalent to optimizing over pure-state
decompositions of the complementary state $\rho_{EB}$.

Let $|\Psi\rangle_{ABE}$ purify $\rho_{AB}$, and write
$\rho_{EB}=\Tr_A|\Psi\rangle\!\langle\Psi|$.
For $X\in\{A,E\}$, let
$\Ens_{X\to B}(\rho_{XB})$ denote the ensembles
$\{p_x,\omega_x\}_x$ that can be prepared on $B$ by measuring $X$.
Outcomes with zero probability are omitted from the ensembles.
The remote assisted gain is
\begin{align}
    \cA_{X\to B}^{\cR}(\rho_{XB})
    =
    \sup_{\{p_x,\omega_x\}\in\Ens_{X\to B}(\rho_{XB})}
    \sum_x p_x\cR(\omega_x)
    -\cR(\rho_B).
    \label{eq:app_assistance_definition}
\end{align}
Here $\cR(\omega_x)$ is the value Bob can obtain from the
conditional input $\omega_x$, while $\cR(\rho_B)$ is the value
available without knowing the measurement outcome.

To express this optimization in terms of pure-state
decompositions, define
\begin{align}
    \widehat{\cR}(\sigma_{XB})
    =
    \sup_{\{p_k,\psi_k\}}
    \sum_k p_k\,
    \cR\!\left(
        \Tr_X|\psi_k\rangle\!\langle\psi_k|
    \right),
    \label{eq:app_upper_roof}
\end{align}
where the supremum runs over all pure-state decompositions
$\sigma_{XB}=\sum_k p_k|\psi_k\rangle\!\langle\psi_k|$.

\begin{proposition}[Complementary roofs]
\label{prop:app_complementary_roofs}
For every convex task score $\cR$, the assisted gains satisfy
\begin{align}
    \cA_{A\to B}^{\cR}(\rho_{AB})
    &=
    \widehat{\cR}(\rho_{EB})-\cR(\rho_B),
    \label{eq:app_alice_complementary_roof}
    \\
    \cA_{E\to B}^{\cR}(\rho_{EB})
    &=
    \widehat{\cR}(\rho_{AB})-\cR(\rho_B).
    \label{eq:app_eve_complementary_roof}
\end{align}
Consequently, the exclusive advantage
$\Delta^{\cR}(\rho_{AB})
=\cA_{A\to B}^{\cR}(\rho_{AB})
-\cA_{E\to B}^{\cR}(\rho_{EB})$
is
\begin{align}
    \Delta^{\cR}(\rho_{AB})
    =
    \widehat{\cR}(\rho_{EB})
    -
    \widehat{\cR}(\rho_{AB}).
    \label{eq:app_delta_roof}
\end{align}
\end{proposition}

\begin{proof}
First, we show that Alice's optimization in
Eq.~\eqref{eq:app_assistance_definition} may be restricted
to rank-one POVMs.
Consider any POVM $\{M_a^A\}_a$ and decompose each effect as
\begin{align}
    M_a^A
    =
    \sum_\mu
    |\alpha_{a\mu}\rangle\!\langle\alpha_{a\mu}|.
\end{align}
The rank-one effects
$\{|\alpha_{a\mu}\rangle\!\langle\alpha_{a\mu}|\}_{a,\mu}$
form a POVM whose outcomes retain both labels $a$ and $\mu$.
Forgetting $\mu$ recovers the original conditional ensemble,
so
\begin{align}
    p_a\omega_a
    =
    \sum_\mu p_{a\mu}\omega_{a\mu},
    \qquad
    p_a=\sum_\mu p_{a\mu}.
\end{align}
For $p_a>0$, the state $\omega_a$ is therefore a mixture of
the states $\omega_{a\mu}$ with probabilities
$p_{a\mu}/p_a$. Convexity of $\cR$ gives
\begin{align}
    p_a\cR(\omega_a)
    \leq
    \sum_\mu p_{a\mu}\cR(\omega_{a\mu}).
    \label{eq:app_rank_one_refinement}
\end{align}
Summing over $a$ shows that retaining the additional label
$\mu$ cannot decrease the quantity being maximized.
Thus rank-one POVMs suffice for Alice's optimization.

We next show that every such POVM produces a decomposition
allowed in Eq.~\eqref{eq:app_upper_roof}, with
$\sigma_{XB}=\rho_{EB}$.
Relabeling the outcomes by $a$, write
$M_a^A=|\alpha_a\rangle\!\langle\alpha_a|$ and define
\begin{align}
    |\chi_a\rangle_{EB}
    =
    (\langle\alpha_a|_A\otimes\mathbb I_{EB})
    |\Psi\rangle_{ABE}.
\end{align}
The probability of outcome $a$ is
$p_a=\langle\chi_a|\chi_a\rangle$.
For $p_a>0$, the conditional state of $EB$ is the pure state
\begin{align}
    |\psi_a\rangle_{EB}
    =
    \frac{|\chi_a\rangle_{EB}}{\sqrt{p_a}}.
\end{align}
Completeness of the POVM,
$\sum_a|\alpha_a\rangle\!\langle\alpha_a|=\mathbb I_A$,
implies
\begin{align}
    \sum_a p_a|\psi_a\rangle\!\langle\psi_a|
    =
    \Tr_A|\Psi\rangle\!\langle\Psi|
    =
    \rho_{EB}.
\end{align}
Thus these conditional pure states form a decomposition of
$\rho_{EB}$. Bob receives their $B$ marginals, $\omega_a=\Tr_E|\psi_a\rangle\!\langle\psi_a|$,
which need not be pure. Consequently,
\begin{align}
    \sum_a p_a\cR(\omega_a)=
    \sum_a p_a\cR\!\left(
        \Tr_E|\psi_a\rangle\!\langle\psi_a|
    \right)\leq
    \widehat{\cR}(\rho_{EB}).
\end{align}
Since rank-one POVMs suffice, this proves
\begin{align}
    \sup_{\{p_a,\omega_a\}\in\Ens_{A\to B}(\rho_{AB})}
    \sum_a p_a\cR(\omega_a)
    \leq
    \widehat{\cR}(\rho_{EB}).
\end{align}

To prove the reverse inequality, consider any pure-state
decomposition of $\rho_{EB}$ entering
Eq.~\eqref{eq:app_upper_roof}.
The HJW theorem~\cite{hughston1993complete} ensures that
Alice can prepare this decomposition by a rank-one POVM
on its purifying system $A$. Bob's conditional states are then precisely the
$B$ marginals appearing in Eq.~\eqref{eq:app_upper_roof}.
Therefore every value entering that supremum also occurs
in Alice's optimization. Combining the two inequalities gives
\begin{align}
    \sup_{\{p_a,\omega_a\}\in\Ens_{A\to B}(\rho_{AB})}
    \sum_a p_a\cR(\omega_a)
    =
    \widehat{\cR}(\rho_{EB}).
\end{align}
Subtracting $\cR(\rho_B)$ gives
Eq.~\eqref{eq:app_alice_complementary_roof}.
Exchanging $A$ and $E$ gives
Eq.~\eqref{eq:app_eve_complementary_roof}.
Finally, subtracting the two assisted gains cancels their
common term $\cR(\rho_B)$ and yields
Eq.~\eqref{eq:app_delta_roof}.
\end{proof}

Convexity is needed only for
Eq.~\eqref{eq:app_rank_one_refinement}, which allows us to
restrict the measurement optimization to rank-one POVMs.
Without convexity, forgetting part of the measurement outcome
can increase $\sum_a p_a\cR(\omega_a)$, so the pure-state
optimization in Eq.~\eqref{eq:app_upper_roof} need not equal
the unrestricted measurement optimization.

\section{B. Robustness under weak noise}
\label{app:robustness}

We show that Alice's advantage at an entangled pure state
persists under small perturbations. We keep her measurement
fixed and use continuity to compare its value with Eve's
optimized value.

\begin{proof}[Proof of Theorem~\ref{thm:robustness}]
Write
\begin{align}
    |\Phi\rangle_{AB}
    =
    \sum_{j=1}^r\sqrt{\lambda_j}\,
    |j\rangle_A|j\rangle_B,
    \qquad r\geq2,\quad \lambda_j>0,
\end{align}
in Schmidt form, and set $r_0=\cR(\Phi_B)$.
Alice's measurement in a basis containing the Schmidt
vectors prepares $|j\rangle_B$ with probability $\lambda_j$.
Since $\Phi_B=\sum_j\lambda_j|j\rangle\!\langle j|_B$,
strict convexity gives
\begin{align}
    \delta
    :=
    \sum_j\lambda_j\cR(|j\rangle\!\langle j|_B)-r_0
    >0.
\end{align}
Eve is uncorrelated with Bob at $\Phi_{AB}$, so her
measurement leaves his state unchanged and her optimized
value is $r_0$.

Keep Alice's measurement fixed. For any state $\rho_{AB}$,
let $p_a$ and $\omega_a$ be its outcome probabilities and
Bob's conditional states, and define
\begin{align}
    G_A(\rho_{AB})=\sum_{a:p_a>0}p_a\cR(\omega_a).
\end{align}
In particular, $G_A(\Phi_{AB})=r_0+\delta$.
To see that $G_A$ is continuous, note that
\begin{align}
    p_a\omega_a
    =
    \Tr_A[(M_a^A\otimes\mathbb I_B)\rho_{AB}]
\end{align}
depends linearly on $\rho_{AB}$. Thus $p_a$, $\omega_a$,
and $p_a\cR(\omega_a)$ are continuous wherever $p_a>0$.
When $p_a\to0$, boundedness of the continuous function
$\cR$ on the compact state space gives
\begin{align}
    |p_a\cR(\omega_a)|
    \leq p_a\max_{\omega_B}|\cR(\omega_B)|
    \longrightarrow0.
\end{align}
Hence $G_A$ is continuous also when some outcomes have
zero probability.

Eve's optimized value is $\widehat{\cR}(\rho_{AB})$ by
Proposition~\ref{prop:app_complementary_roofs}.
The function
$|\psi\rangle\!\langle\psi|\mapsto
\cR(\Tr_A|\psi\rangle\!\langle\psi|)$
is continuous and bounded on pure states.
Its upper roof is therefore continuous by
Corollary~2 of Ref.~\cite{shirokov2010properties}.
Since every pure-state decomposition of $\Phi_{AB}$
consists of the same projector,
$\widehat{\cR}(\Phi_{AB})=\cR(\Phi_B)=r_0$.

Consequently,
\begin{align}
    G_A(\rho_{AB})-\widehat{\cR}(\rho_{AB})
    \longrightarrow\delta>0
    \qquad\text{as }\rho_{AB}\longrightarrow\Phi_{AB}.
\end{align}
There is therefore $\epsilon^*>0$ such that this
difference is at least $\delta/2$ whenever
$\|\rho_{AB}-\Phi_{AB}\|_1\leq\epsilon^*$.
Alice's optimized value is at least $G_A(\rho_{AB})$,
whereas Eve's equals $\widehat{\cR}(\rho_{AB})$.
The common term $\cR(\rho_B)$ cancels, so
\begin{align}
    \Delta^{\cR}(\rho_{AB})
    \geq G_A(\rho_{AB})-\widehat{\cR}(\rho_{AB})
    \geq\frac{\delta}{2}>0.
\end{align}
\end{proof}

\section{D. Spectral resources and entanglement roofs}
\label{app:spectral_roofs}

We show that, for a convex spectral task score, the exclusive
advantage equals the difference of an entanglement monotone
evaluated on $\rho_{AB}$ and $\rho_{EB}$.

Let $\cR$ be convex and spectral, meaning unitarily invariant.
All pure states of $B$ then have the same value,
\begin{align}
    \cR_{\rm pure}
    =
    \cR(|b\rangle\!\langle b|).
\end{align}
For a bipartite pure state $|\phi\rangle_{XB}$, define
\begin{align}
    e_{\cR}(|\phi\rangle_{XB})
    =
    \cR_{\rm pure}-\cR(\phi_B),
    \qquad
    \phi_B=\Tr_X|\phi\rangle\!\langle\phi|.
    \label{eq:app_induced_pure_entanglement}
\end{align}
Every state of $B$ is a mixture of pure states, so convexity
gives $\cR(\omega_B)\leq\cR_{\rm pure}$.
Thus $e_{\cR}$ is nonnegative and vanishes on pure product
states. Since $\cR_{\rm pure}-\cR(\omega_B)$ is concave
and unitarily invariant, Vidal's criterion
\cite{vidal2000monotones} implies that $e_{\cR}$ is a pure-state
entanglement monotone. Its convex-roof extension,
\begin{align}
    E_{\cR}(\sigma_{XB})
    =
    \inf_{\{q_k,\phi_k\}}
    \sum_k q_k e_{\cR}(|\phi_k\rangle_{XB}),
    \label{eq:app_induced_entanglement_roof}
\end{align}
is also an entanglement monotone. 

Because $\cR_{\rm pure}$ is constant, maximizing the average
of $\cR(\phi_B)$ is equivalent to minimizing that of
$e_{\cR}(|\phi\rangle)$ over the same decompositions:
\begin{align}
    \widehat{\cR}(\sigma_{XB})
=
    \sup_{\{q_k,\phi_k\}}
    \sum_kq_k
    \bigl[\cR_{\rm pure}-e_{\cR}(|\phi_k\rangle)\bigr]
=
    \cR_{\rm pure}-E_{\cR}(\sigma_{XB}).
    \label{eq:app_roof_complementarity}
\end{align}
Substituting into Eq.~\eqref{eq:app_delta_roof} gives
\begin{align}
    \Delta^{\cR}(\rho_{AB})
    =
    E_{\cR}(\rho_{AB})-E_{\cR}(\rho_{EB}).
    \label{eq:app_spectral_delta}
\end{align}
Thus Alice has a strict exclusive advantage precisely when
$E_{\cR}(\rho_{AB})>E_{\cR}(\rho_{EB})$, as stated in the
main text.

For the purity functional
\begin{align}
    \mathcal P(\omega_B)
    =
    \log d_B-S(\omega_B),
\end{align}
we have $\mathcal P_{\rm pure}=\log d_B$, so
$e_{\mathcal P}(|\phi\rangle)=S(\phi_B)$.
Its convex roof is therefore the entanglement of formation:
\begin{align}
    E_{\mathcal P}(\sigma_{XB})
    =
    E_F(\sigma_{XB})
    =
    \inf_{\{q_k,\phi_k\}}
    \sum_kq_k S(\phi_B^k).
\end{align}
Equation~\eqref{eq:app_spectral_delta} then becomes
\begin{align}
    \Delta^{\mathcal P}(\rho_{AB})
    =
    E_F(\rho_{AB})-E_F(\rho_{EB}).
    \label{eq:app_purity_delta}
\end{align}
The same identity follows from the Koashi--Winter relation~\cite{koashi2004monogamy}.

\section{E. Work extraction and machine dependence}
\label{app:work_extraction}

We derive the exclusive advantages for bath-assisted work
and ergotropy, explain when their entanglement roofs can
be evaluated using concurrence, and construct a state
for which the two tasks favor different parties.

\subsection{Work extraction in a bath}

We use the work value
$\mathcal W_\beta(\omega)=k_BT D(\omega\Vert\gamma_B^\beta)$,
where $\gamma_B^\beta=e^{-\beta H_B}/Z_B$,
$Z_B=\Tr e^{-\beta H_B}$, and $\beta=(k_BT)^{-1}$ with $T>0$.
Entropies and relative entropies use natural logarithms.
Expanding gives
\begin{align}
    \mathcal W_\beta(\omega)
    =
    \Tr(H_B\omega)-k_BT S(\omega)+k_BT\ln Z_B.
\end{align}
For every ensemble $\rho_B=\sum_a t_a\tau_a$,
\begin{align}
    \sum_a t_a\mathcal W_\beta(\tau_a)
    =
    \Tr(H_B\rho_B)+k_BT\ln Z_B -k_BT\sum_a t_aS(\tau_a).
    \label{eq:app_average_free_energy}
\end{align}
The first two terms are fixed, so maximizing the average
work amounts to minimizing $\sum_a t_aS(\tau_a)$.
By the rank-one refinement and HJW argument of
Appendix~\ref{app:complementary_roofs},
\begin{align}
    \inf_{\{t_a,\tau_a\}\in\Ens_{A\to B}(\rho_{AB})}
    \sum_a t_aS(\tau_a)
    =
    E_F(\rho_{EB}).
\end{align}
Subtracting $\mathcal W_\beta(\rho_B)$ gives Alice's
assisted gain, and exchanging $A$ and $E$ gives Eve's:
\begin{align}
    \cA_{A\to B}^{\mathcal W_\beta}
    &=
    k_BT\,[S(\rho_B)-E_F(\rho_{EB})],
    \nonumber\\
    \cA_{E\to B}^{\mathcal W_\beta}
    &=
    k_BT\,[S(\rho_B)-E_F(\rho_{AB})].
    \label{eq:app_work_assistances}
\end{align}
Consequently,
\begin{align}
    \Delta^{\mathcal W_\beta}(\rho_{AB})
    =
    k_BT\,[E_F(\rho_{AB})-E_F(\rho_{EB})].
    \label{eq:app_work_delta}
\end{align}
This signed difference is positive precisely when Alice
has an exclusive advantage in bath-assisted work extraction.

These formulas concern the one-copy remote-preparation
model defined in the main text. If collective measurements
on $A^n$ or $E^n$ are allowed, the corresponding asymptotic
per-copy formulas instead involve the regularized
entanglement of formation. Assisted work distillation
and this regularization are discussed in
Ref.~\cite{morris2019assisted}.

\subsection{Ergotropic work}

Let the ground energy of $H_B$ be zero and define
the passive energy
\begin{align}
    E_{\rm p}^{H_B}(\omega)
    =
    \min_U\Tr(H_BU\omega U^\dagger).
    \label{eq:app_passive_energy_variational}
\end{align}
This function is concave, being a pointwise minimum of
affine functions, and unitarily invariant. It vanishes
on pure states because a unitary can map any pure state
to a ground state.
By the criterion used in Appendix~\ref{app:spectral_roofs},
$E_{\rm p}^{H_B}(\phi_B)$ is therefore a pure-state
entanglement monotone. Denote its convex roof by
\begin{align}
    \mathcal E_{\rm p}(\sigma_{XB})
    =
    \inf_{\{q_k,\phi_k\}}
    \sum_k q_k E_{\rm p}^{H_B}(\phi_B^k),
    \label{eq:app_passive_roof}
\end{align}
where the infimum runs over pure-state decompositions
of $\sigma_{XB}$.
These properties of passive energy are also established
in Ref.~\cite{Alimuddin2020}.

Ergotropy is
\begin{align}
    \mathcal W_{\rm erg}(\omega)
    =
    \Tr(H_B\omega)-E_{\rm p}^{H_B}(\omega).
\end{align}
Every remotely prepared ensemble has the same average
energy $\Tr(H_B\rho_B)$, so maximizing its average
ergotropy amounts to minimizing its average passive energy.
Concavity allows restriction to rank-one POVMs, and HJW
identifies this minimum with the complementary convex roof.
Subtracting the unassisted ergotropy gives
\begin{align}
    \cA_{A\to B}^{\mathcal W_{\rm erg}}
    &=
    E_{\rm p}^{H_B}(\rho_B)
    -\mathcal E_{\rm p}(\rho_{EB}),
    \nonumber\\
    \cA_{E\to B}^{\mathcal W_{\rm erg}}
    &=
    E_{\rm p}^{H_B}(\rho_B)
    -\mathcal E_{\rm p}(\rho_{AB}).
\end{align}
Thus
\begin{align}
    \Delta^{\mathcal W_{\rm erg}}(\rho_{AB})
    =
    \mathcal E_{\rm p}(\rho_{AB})
    -\mathcal E_{\rm p}(\rho_{EB}).
    \label{eq:app_ergotropy_delta}
\end{align}
Its positive part,
$\max\{0,\Delta^{\mathcal W_{\rm erg}}(\rho_{AB})\}$,
is the exclusive ergotropic work.
The measurement-assisted optimization is the general-POVM
version of daemonic ergotropy, originally introduced
for projective measurements in
Ref.~\cite{francica2017daemonic}.

\subsection{The two-qubit reduction}
\label{app:two_qubit_reduction}

Let $\sigma_{XB}$ be a two-qubit state.
For a pure state $|\phi\rangle_{XB}$, the eigenvalues
of $\phi_B$ are
\begin{align}
    \lambda_\pm(C)
    =
    \frac{1\pm\sqrt{1-C^2}}{2},
    \label{eq:app_schmidt_concurrence}
\end{align}
where $C$ is its concurrence.
Thus every pure-state spectral functional is a function
of $C$, but this alone does not determine its mixed-state
convex roof.

Suppose
$e(|\phi\rangle)=f(C(|\phi\rangle))$,
where $f$ is convex and nondecreasing.
For every decomposition of $\sigma$,
\begin{align}
    \sum_kq_k f(C_k)
    \geq
    f\!\left(\sum_kq_kC_k\right)
    \geq
    f(C(\sigma)).
    \label{eq:app_concurrence_lower_bound}
\end{align}
The first inequality uses convexity of $f$.
The second uses its monotonicity and the fact that
$C(\sigma)$ is the convex roof of pure-state concurrence.
Wootters constructs a decomposition in which every
component has concurrence $C(\sigma)$
\cite{wootters1998}. This decomposition attains the bound,
so the convex roof of $e$ is
\begin{align}
    E(\sigma)=f(C(\sigma)).
    \label{eq:app_flat_roof_formula}
\end{align}

Both the entropy of entanglement and the passive energy
of a qubit are convex, nondecreasing functions of
pure-state concurrence. In particular, for
$H_B=\varepsilon|1\rangle\!\langle1|$ with $\varepsilon>0$,
the passive energy is $\varepsilon\lambda_-(C)$, giving
\begin{align}
    \mathcal E_{\rm p}(\sigma)
    =
    \frac{\varepsilon}{2}
    \left(1-\sqrt{1-C(\sigma)^2}\right).
    \label{eq:app_qubit_passive_roof}
\end{align}

Without convexity of $f$, the convex roof need not be
determined by concurrence alone.
To see this, consider $g(C)=\sqrt C$ and the state
\begin{align}
    \rho_c
    =
    c|\Phi^+\rangle\!\langle\Phi^+|
    +(1-c)|00\rangle\!\langle00|,
    \qquad 0<c<1,
\end{align}
whose concurrence is $C(\rho_c)=c$.
The two components have concurrence $1$ and $0$, so this
decomposition gives $E_g(\rho_c)\leq c$.
Conversely, every pure-state decomposition satisfies
\begin{align}
    \sum_kq_k\sqrt{C_k}
    \geq
    \sum_kq_kC_k
    \geq
    C(\rho_c)=c,
\end{align}
because $\sqrt{C_k}\geq C_k$ and concurrence is itself
a convex roof. Hence $E_g(\rho_c)=c$.
A pure state with the same concurrence instead has
$E_g=\sqrt c>c$, showing that equal concurrence does
not imply equal values of this roof.

The dimensions of the complementary state also matter.
If $A$ and $B$ are qubits, a minimal purification has
$\dim E=\operatorname{rank}\rho_{AB}$, which can be four.
Wootters' formula evaluates both complementary roofs
when $\operatorname{rank}\rho_{AB}\leq2$, because both
states are then supported on at most two qubits.
For a full-rank two-qubit $\rho_{AB}$, it evaluates
the $A{:}B$ term but does not directly evaluate
the $E{:}B$ term.

\subsection{An exact cross-task reversal}
\label{app:task_reversal}

We construct a state for which Alice has an exclusive
advantage in ergotropic work extraction, while Eve
outperforms her in purification and bath-assisted work.
Let $B=B_0\oplus B_1$, with $\dim B_0=\dim B_1=3$, and set
\begin{align}
    \lambda=\left(\frac12,\frac12,0\right),
    \qquad
    \mu=\left(\frac45,\frac1{10},\frac1{10}\right).
\end{align}
Take $A$ and $E$ to be four-dimensional, each with
orthonormal basis
$\{|0\rangle,|1\rangle,|2\rangle,|\perp\rangle\}$. Define
\begin{align}
    |\Omega\rangle_{ABE}
    =
    \sqrt{\frac37}
    \sum_{j=0}^2\sqrt{\lambda_j}\,
    |j\rangle_A|j\rangle_{B_0}|\perp\rangle_E +
    \sqrt{\frac47}
    \sum_{j=0}^2\sqrt{\mu_j}\,
    |\perp\rangle_A|j\rangle_{B_1}|j\rangle_E.
    \label{eq:app_task_reversal_state}
\end{align}
Here $|j\rangle_{B_0}=|j\rangle_B$ and
$|j\rangle_{B_1}=|j+3\rangle_B$.

Let $\rho_{AB}=\Tr_E|\Omega\rangle\!\langle\Omega|$
and $\rho_{EB}=\Tr_A|\Omega\rangle\!\langle\Omega|$.
Since $\langle\perp|j\rangle=0$, tracing out either
$A$ or $E$ removes the cross terms between the two branches.
Each reduced state is therefore a mixture of branches
that Bob can distinguish by measuring whether his system
lies in $B_0$ or $B_1$.
In the first branch, Alice and Bob share a pure state
with Schmidt spectrum $\lambda$, while Eve is uncorrelated.
In the second, Eve and Bob share a pure state with
Schmidt spectrum $\mu$, while Alice is uncorrelated.

For either $E_F$ or $\mathcal E_{\rm p}$, convexity bounds
the value of each reduced state from above by the weighted
values of its branches. Bob's local measurement distinguishing
the branches gives the reverse inequality, because
entanglement cannot increase on average.
The value therefore equals this weighted average,
with the product branch contributing zero.

Writing $H_2(\nu)=-\sum_j\nu_j\log_2\nu_j$, we obtain
\begin{align}
    \frac{\Delta^{\mathcal P}}{\ln2}
    =
    \frac37H_2(\lambda)-\frac47H_2(\mu)
    \simeq-0.098245.
    \label{eq:app_task_reversal_purity}
\end{align}
The factor $\ln2$ converts the entropies in bits to our
natural-logarithm convention.
Eve therefore outperforms Alice in purification and,
by Eq.~\eqref{eq:app_work_delta}, bath-assisted work.

For ergotropy, choose
\begin{align}
    H_B
    =
    \epsilon\sum_{j=0}^{5}j|j\rangle\!\langle j|,
    \qquad \epsilon>0.
\end{align}
Passive energy allows all unitaries on $B$, so the
eigenvalues are assigned to the lowest energies regardless
of their original sector.
Padding $\lambda$ and $\mu$ with zeros to dimension six
therefore gives
$E_{\rm p}^{H_B}(\lambda)=\epsilon/2$ and
$E_{\rm p}^{H_B}(\mu)=3\epsilon/10$.
Hence
\begin{align}
    \Delta^{\mathcal W_{\rm erg}}
    =
    \frac37\frac{\epsilon}{2}
    -
    \frac47\frac{3\epsilon}{10}
    =
    \frac{3\epsilon}{70}>0.
    \label{eq:app_task_reversal_ergotropy}
\end{align}
Alice thus outperforms Eve in ergotropic work extraction.

\section{F. Depolarized Bell states and coherence}

We derive the spectral thresholds displayed in Fig.~2,
identify an entangled region where Eve can simulate every
ensemble available to Alice, and compare these results
with the coherence threshold.

\subsection{Spectral roofs and the task-independent obstruction}
\label{app:bell_diagonal}
\label{app:entanglement_not_sufficient}

For a two-qubit Bell-diagonal state
\begin{align}
    \rho_{AB}
    =
    \frac14\left(
        \mathbb I\otimes\mathbb I
        +\sum_{i=1}^3c_i\sigma_i\otimes\sigma_i
    \right),
\end{align}
define
\begin{align}
    \kappa=\max_i|c_i|,
    \qquad
    C=\max\{0,2\lambda_{\max}(\rho_{AB})-1\}.
\end{align}
Since $\rho_B=\mathbb I/2$, the classical correlation
optimized over Alice's measurements is
$J(B|A)=\ln2-h[(1+\kappa)/2]$, where
$h(x)=-x\ln x-(1-x)\ln(1-x)$
\cite{luo2008quantum}.
The Koashi--Winter relation gives
$E_F(\rho_{EB})=S(\rho_B)-J(B|A)$.
Together with Wootters' formula, this yields
\begin{align}
    E_F(\rho_{EB})
    &=
    h\!\left(\frac{1+\kappa}{2}\right),
    \nonumber\\
    \Delta^{\mathcal P}
    &=
    h\!\left(\frac{1+\sqrt{1-C^2}}{2}\right)
    -
    h\!\left(\frac{1+\kappa}{2}\right).
    \label{eq:app_bell_delta}
\end{align}
Because $h$ is strictly decreasing on $[1/2,1]$,
\begin{align}
    \Delta^{\mathcal P}>0
    \quad\Longleftrightarrow\quad
    C^2+\kappa^2>1.
    \label{eq:app_bell_criterion}
\end{align}

Consider now the depolarized Bell family
\begin{align}
    \rho_{AB}(p)
    =
    (1-p)|\Phi^+\rangle\!\langle\Phi^+|
    +p\frac{\mathbb I_{AB}}4,
    \qquad 0\leq p\leq1.
    \label{eq:app_depolarizing_choi}
\end{align}
Here $\kappa=1-p$ and
$C_{AB}=\max\{0,1-3p/2\}$.
Every nonzero outcome of a rank-one POVM on $A$ prepares
a Bob state with spectrum $(1-p/2,p/2)$, independently
of the measurement direction.
By HJW, every pure-state decomposition of $\rho_{EB}$
therefore has this same Bob-side spectrum. In particular,
\begin{align}
    E_F(\rho_{EB})=h(p/2).
\end{align}

More generally, suppose a spectral task induces a
pure-state entanglement functional $f_{\cR}(C)$
that is convex and strictly increasing.
The two-qubit reduction of
Appendix~\ref{app:two_qubit_reduction} and the fixed
spectrum above give
\begin{align}
    E_{\cR}(\rho_{AB})
    &=f_{\cR}(C_{AB}),
    \nonumber\\
    E_{\cR}(\rho_{EB})
    &=f_{\cR}(C_E),
    \qquad
    C_E=\sqrt{2p-p^2}.
\end{align}
Here $C_E$ parametrizes the common Schmidt spectrum
of the pure-state components of $\rho_{EB}$; it is
not a mixed-state two-qubit concurrence.
Thus the advantage is positive exactly when $C_{AB}>C_E$,
or equivalently
\begin{align}
    0\leq p<p_\star,
    \qquad
    p_\star
    =
    \frac{10-4\sqrt3}{13}
    \simeq0.236292.
    \label{eq:app_spectral_threshold}
\end{align}
The advantages vanish at $p=p_\star$ and are negative
for $p>p_\star$. The convexity assumption on $f_{\cR}$
is needed to evaluate the roof of $\rho_{AB}$.

The same family contains an entangled region where
Eve can simulate every ensemble Alice can prepare.
If a CPTP map $\Lambda_{E\to A}$ satisfies
\begin{align}
    \rho_{AB}
    =
    (\Lambda_{E\to A}\otimes\id_B)(\rho_{EB}),
    \label{eq:app_antidegradable_state}
\end{align}
then $\Lambda_{E\to A}^\dagger$ maps every Alice POVM
to an Eve POVM with the same conditional states on $B$.
Hence
\begin{align}
    \Ens_{A\to B}(\rho_{AB})
    \subseteq
    \Ens_{E\to B}(\rho_{EB}).
\end{align}
The state $\rho_{AB}(p)$ is the normalized Choi state
of the depolarizing channel
$\mathcal N_p(\omega)=(1-p)\omega+p\mathbb I/2$.
This channel is antidegradable exactly for $p\geq1/3$
\cite{paddock2017characterization}, giving
Eq.~\eqref{eq:app_antidegradable_state} in that range. Meanwhile, the smallest eigenvalue of
$\rho_{AB}(p)^{T_B}$ is $-1/2+3p/4$, so the state
is entangled exactly for $p<2/3$. 

The parameter regions can therefore be summarized as
\begin{align}
\begin{array}{c@{\quad}l}
0\leq p<p_\star
    & \text{strict advantage for the spectral tasks above},\\
p=p_\star
    & \text{their exclusive advantages vanish},\\
p_\star<p<1/3
    & \text{their exclusive advantages are negative},\\
1/3\leq p<2/3
    & \text{entangled, but Eve reproduces every Alice ensemble},\\
2/3\leq p\leq1
    & \text{separable}.
\end{array}
\label{eq:app_three_regions}
\end{align}

\subsection{Coherence: exact value and analytic bound}
\label{app:coherence}

Relative-entropy coherence provides the nonspectral comparison in
Fig.~2. Fix the computational basis of $B$ as the
incoherent basis, and let $\Delta_B$ denote dephasing
in that basis. Relative-entropy coherence is
\begin{align}
    C_{\rm r}(\omega)
    =
    S(\Delta_B\omega)-S(\omega).
\end{align}
Its convexity gives
\begin{align}
    \Delta^{C_{\rm r}}(\rho_{AB})
    =
    \widehat C_{\rm r}(\rho_{EB})
    -
    \widehat C_{\rm r}(\rho_{AB}),
    \label{eq:app_coherence_delta}
\end{align}
where
\begin{align}
    \widehat C_{\rm r}(\sigma_{XB})
    =
    \sup_{\{q_k,\phi_k\}}
    \sum_kq_k
    \left[
        S(\Delta_B\phi_B^k)-S(\phi_B^k)
    \right].
    \label{eq:app_coherence_roof}
\end{align}
The supremum runs over pure-state decompositions
of $\sigma_{XB}$, with
$\phi_B^k=\Tr_X|\phi_k\rangle\!\langle\phi_k|$.

For work extraction, the average energy is fixed by $\rho_B$
and cancels from the advantage, leaving only spectral terms.
For coherence, $S(\Delta_B\phi_B^k)$ depends on the conditional
state's orientation relative to the incoherent basis, and its
average is not fixed by $\rho_B$:
\begin{align}
    \sum_kq_kS(\Delta_B\phi_B^k)
    \neq S(\Delta_B\rho_B)
    \qquad\text{in general}.
\end{align}
The two terms in
$C_{\rm r}(\phi_B^k)=S(\Delta_B\phi_B^k)-S(\phi_B^k)$
must therefore be optimized over the same decomposition.
Thus $\widehat C_{\rm r}$ need not reduce to a
basis-independent entanglement roof, and the coherence
advantage can have a different threshold even when $B$
is a qubit.

\medskip
\noindent
{\it Coherence threshold.--}
We now use the family
\begin{align}
    \rho_{AB}(p)
    =
    (1-p)|\Phi^+\rangle\!\langle\Phi^+|
    +p\frac{\mathbb I_{AB}}4,
    \qquad 0\leq p\leq1,
    \label{eq:app_coherence_depolarized}
\end{align}
and denote its complementary state by $\rho_{EB}(p)$. We take the
computational basis as the incoherent basis and write
\begin{align}
h_2(x)
=
-x\log_2x-(1-x)\log_2(1-x).
\end{align}
The reduced state is $\rho_B=\mathbb I_B/2$ and therefore
$C_{\rm r}(\rho_B)=0$.

For a qubit state
$\omega=(\mathbb I+\boldsymbol r\cdot\boldsymbol\sigma)/2$,
\begin{align}
C_{\rm r}(\omega)
=
h_2\!\left(\frac{1+r_z}{2}\right)
-
h_2\!\left(\frac{1+|\boldsymbol r|}{2}\right).
\label{eq:app_qubit_coherence}
\end{align}
Every state remotely prepared by measuring $A$ has a Bloch vector
satisfying $|\boldsymbol r|\leq1-p$. Consequently,
\begin{align}
C_{\rm r}(\omega)
\leq
1-h_2\!\left(\frac{p}{2}\right).
\end{align}
The bound is attained by measuring $A$ in any equatorial basis,
for example the eigenbasis of $\sigma_x$. The two conditional
states on $B$ are then
\begin{align}
\tau_{\pm}
=
\frac{1}{2}
\left[
\mathbb I_B
\pm(1-p)\sigma_x
\right],
\end{align}
and each has coherence $1-h_2(p/2)$. It follows that
\begin{align}
\widehat C_{\rm r}\!\left(\rho_{EB}(p)\right)
=
1-h_2\!\left(\frac{p}{2}\right).
\label{eq:app_alice_coherence}
\end{align}

We next bound Eve's coherence roof. For every pure-state
decomposition
$\rho_{AB}(p)=\sum_kq_k|\phi_k\rangle\langle\phi_k|$,
\begin{align}
\sum_kq_k C_{\rm r}(\phi_B^k)
&\leq
1-\sum_kq_kS(\phi_B^k)
\nonumber\\
&\leq
1-E_F\!\left(\rho_{AB}(p)\right),
\end{align}
where the first inequality follows from
$S(\Delta_B\phi_B^k)\leq1$, and the second follows from the
definition of the entanglement of formation. Hence,
\begin{align}
\widehat C_{\rm r}\!\left(\rho_{AB}(p)\right)
\leq
1-E_F\!\left(\rho_{AB}(p)\right).
\label{eq:app_eve_coherence_bound}
\end{align}
Combining Eqs.~\eqref{eq:app_coherence_delta},
\eqref{eq:app_alice_coherence}, and
\eqref{eq:app_eve_coherence_bound} gives
\begin{align}
\Delta^{C_{\rm r}}\!\left(\rho_{AB}(p)\right)
\geq
E_F\!\left(\rho_{AB}(p)\right)
-
h_2\!\left(\frac{p}{2}\right).
\label{eq:app_coherence_lower_bound}
\end{align}

Since $\rho_{AB}(p)$ is Bell diagonal, its concurrence is
\begin{align}
\mathcal C(p)
=
\max\left\{
0,1-\frac{3p}{2}
\right\},
\end{align}
and Wootters' formula gives~\cite{wootters1998}
\begin{align}
E_F\!\left(\rho_{AB}(p)\right)
=
h_2\!\left(
\frac{1+\sqrt{1-\mathcal C(p)^2}}{2}
\right).
\label{eq:app_depolarized_eof}
\end{align}
The right-hand side of
Eq.~\eqref{eq:app_coherence_lower_bound} is strictly positive
whenever
\begin{align}
h_2\!\left(
\frac{1+\sqrt{1-(1-3p/2)^2}}{2}
\right)
>
h_2\!\left(\frac{p}{2}\right).
\end{align}
At equality, symmetry and strict monotonicity of $h_2$ on
$[0,1/2]$ imply
\begin{align}
\sqrt{1-\left(1-\frac{3p}{2}\right)^2}
=
1-p,
\end{align}
or equivalently
\begin{align}
13p^2-20p+4=0.
\end{align}
The only root in the relevant parameter range is
\begin{align}
p_{\rm coh}^{\rm cert}
=
\frac{10-4\sqrt{3}}{13}
\simeq
0.236292.
\label{eq:app_coherence_certified_threshold}
\end{align}
We therefore obtain the analytic certification
\begin{align}
0\leq p<p_{\rm coh}^{\rm cert}
\quad\Longrightarrow\quad
\Delta^{C_{\rm r}}\!\left(\rho_{AB}(p)\right)>0.
\label{eq:app_coherence_certified_region}
\end{align}
This condition is sufficient but not necessary, because
Eq.~\eqref{eq:app_eve_coherence_bound} optimizes the two entropy
terms independently and is therefore not generally tight.

The full task-specific boundary can be obtained by evaluating
the roof in Eq.~\eqref{eq:app_coherence_roof}. Introduce the Bell
basis
\begin{align}
|\beta_\mu\rangle
=
(\mathbb I\otimes\sigma_\mu)|\Phi^+\rangle,
\qquad \mu=0,1,2,3,
\end{align}
and write
\begin{align}
\boldsymbol\lambda(p)
=
\left(
1-\frac{3p}{4},
\frac p4,
\frac p4,
\frac p4
\right)
\end{align}
for the Bell spectrum of $\rho_{AB}(p)$. For a probability
vector $\boldsymbol x=(x_0,x_1,x_2,x_3)$, define
\begin{align}
g(\boldsymbol x)
=
\max_{\boldsymbol\theta}
C_{\rm r}\!\left[
\operatorname{Tr}_A
|\psi(\boldsymbol x,\boldsymbol\theta)\rangle
\langle\psi(\boldsymbol x,\boldsymbol\theta)|
\right],
\end{align}
where
\begin{align}
|\psi(\boldsymbol x,\boldsymbol\theta)\rangle
=
\sum_{\mu=0}^3
\sqrt{x_\mu}\,e^{i\theta_\mu}
|\beta_\mu\rangle .
\end{align}
Twirling each pure state over the correlated Pauli operators
$\{\sigma_\nu^*\otimes\sigma_\nu\}_{\nu=0}^3$ leaves its
Bob-side coherence unchanged, since it acts on $B$ as an
incoherent Pauli conjugation, and produces a Bell-diagonal state
with spectrum $\boldsymbol x$. Hence every decomposition of
$\rho_{AB}(p)$ yields a convex decomposition of
$\boldsymbol\lambda(p)$ whose average value of $g$ is at least the
average coherence. Conversely, for any
$\sum_jw_j\boldsymbol x^{\,j}=\boldsymbol\lambda(p)$, the four
twirled copies of each optimal
$|\psi(\boldsymbol x^{\,j},\boldsymbol\theta^{\,j})\rangle$, taken
with weights $w_j/4$, form a valid pure-state decomposition of
$\rho_{AB}(p)$ that attains $\sum_jw_jg(\boldsymbol x^{\,j})$.
Therefore,
\begin{align}
\widehat C_{\rm r}\!\left(\rho_{AB}(p)\right)
=
\operatorname{cav}g\!\left(\boldsymbol\lambda(p)\right),
\label{eq:app_coherence_concave_envelope}
\end{align}
where $\operatorname{cav}g$ denotes the upper concave envelope of
$g$. Equivalently,
\begin{align}
\operatorname{cav}g(\boldsymbol\lambda)
=
\max_{\{w_j,\boldsymbol x^{\,j}\}}
\left\{
\sum_jw_jg(\boldsymbol x^{\,j})
:
\sum_jw_j\boldsymbol x^{\,j}
=
\boldsymbol\lambda
\right\},
\label{eq:app_coherence_primal}
\end{align}
where $\{w_j\}$ is a probability distribution and each
$\boldsymbol x^{\,j}$ is a probability vector. Its dual form is
\begin{align}
\operatorname{cav}g(\boldsymbol\lambda)
=
\min_{\boldsymbol y}
\left\{
\boldsymbol y\cdot\boldsymbol\lambda
:
\boldsymbol y\cdot\boldsymbol x
\geq g(\boldsymbol x)
\ \text{for all }\boldsymbol x
\right\}.
\label{eq:app_coherence_dual}
\end{align}
We evaluate $\operatorname{cav}g$ by a cutting-plane scheme on the
dual~\eqref{eq:app_coherence_dual}, iteratively adding the pure
state that maximally violates the current constraint. Any explicit
decomposition~\eqref{eq:app_coherence_primal}, reconstructed and
verified as a decomposition of $\rho_{AB}(p)$, gives a lower bound,
whereas any $\boldsymbol y$ gives the upper bound
$\boldsymbol y\cdot\boldsymbol\lambda
+\max_{\psi}\left[C_{\rm r}(\psi_B)
-\boldsymbol y\cdot\boldsymbol x(\psi)\right]$, with the maximum
taken over all pure two-qubit states. The two bounds agree to
within $10^{-7}$ and give a unique crossing at
\begin{align}
p_{\rm coh}^{\rm num}
\simeq
0.27824.
\label{eq:app_coherence_numerical_threshold}
\end{align}
In particular, for $p>p_{\rm coh}^{\rm num}$ the negativity of
$\Delta^{C_{\rm r}}$ is certified by an explicit decomposition, for
example $\Delta^{C_{\rm r}}(\rho_{AB}(0.27845))
\leq-5.7\times10^{-4}$.
Thus, subject to the stated numerical precision,
\begin{subequations}
\begin{align}
\Delta^{C_{\rm r}}\!\left(\rho_{AB}(p)\right)&>0,
&&0\leq p<p_{\rm coh}^{\rm num},\\
\Delta^{C_{\rm r}}\!\left(\rho_{AB}(p)\right)&=0,
&&p=p_{\rm coh}^{\rm num},\\
\Delta^{C_{\rm r}}\!\left(\rho_{AB}(p)\right)&<0,
&&p_{\rm coh}^{\rm num}<p\leq1.
\end{align}
\end{subequations}
In particular, the interval
$p_{\rm coh}^{\rm cert}\leq p<p_{\rm coh}^{\rm num}$ exhibits
exclusive coherence that is not detected by the
entanglement-of-formation bound.

\section{G. Ensemble simulation and statistical morphisms}
\label{app:simulation}

This section proves the exact task-independent simulation theorem stated in the main text. We first assume that
$\rho_B$ is full rank; the rank-deficient case is reduced to this one
at the end of the section.

Let $R\simeq B$, write
$\rho_B=\sum_j\lambda_j|j\rangle\!\langle j|$, and introduce the
canonical purification
\begin{align}
    |\psi_\rho\rangle_{RB}
    =
    \sum_j\sqrt{\lambda_j}|j\rangle_R|j\rangle_B.
\end{align}
There is an isometry $V:R\to A\otimes E$ such that
\begin{align}
    |\Psi\rangle_{ABE}
    =
    (V\otimes\mathbb I_B)|\psi_\rho\rangle_{RB}.
\end{align}
Define the complementary channels
\begin{align}
    \mathcal N_A(X)
    &={}
    \Tr_E[VXV^\dagger],
    &
    \mathcal N_E(X)
    &={}
    \Tr_A[VXV^\dagger].
    \label{eq:app_complementary_channels}
\end{align}
Then
\begin{align}
    \rho_{AB}
    &={}
    (\mathcal N_A\otimes\id_B)(\psi_\rho),
    &
    \rho_{EB}
    &={}
    (\mathcal N_E\otimes\id_B)(\psi_\rho),
\end{align}
where $\psi_\rho=|\psi_\rho\rangle\!\langle\psi_\rho|$. For every
operator $M_X$ on $X=A,E$,
\begin{align}
    \Tr_X[(M_X\otimes\mathbb I_B)\rho_{XB}]
    =
    \sqrt{\rho_B}\,
    \mathcal N_X^\dagger(M_X)^T
    \sqrt{\rho_B},
    \label{eq:app_faithful_steering_identity}
\end{align}
where the transpose is taken in the eigenbasis used above. Since
$\rho_B$ is full rank, the right-hand side is an injective function of
$\mathcal N_X^\dagger(M_X)$.

A linear map
$\mathcal L:\cL(A)\to\cL(E)$ is a statistical morphism of
$\mathcal N_A$ if, for every POVM $\{F_x\}_x$ on $E$, there is a POVM
$\{G_x\}_x$ on $A$ such that
\begin{align}
    \Tr[\mathcal L(\mathcal N_A(\omega))F_x]
    =
    \Tr[\mathcal N_A(\omega)G_x]
    \label{eq:app_statistical_morphism}
\end{align}
for every input state $\omega$ and every $x$. In the theorem below,
$\mathcal L$ is additionally required to be Hermiticity preserving and
trace preserving on the full algebra. This is the trace-preserving
statistical-morphism notion used in quantum comparison of experiments
\cite{buscemi2012comparison,buscemi2016degradable}.

\begin{theorem}[Exact ensemble-simulation order]
\label{thm:app_statistical_simulation}
The following statements are equivalent:
\begin{enumerate}
    \item
    $\Ens_{E\to B}(\rho_{EB})
    \subseteq\Ens_{A\to B}(\rho_{AB})$.

    \item
    For every POVM $\{F_x\}_x$ on $E$, there is a POVM
    $\{G_x\}_x$ on $A$ such that
    \begin{align}
        \mathcal N_E^\dagger(F_x)
        =
        \mathcal N_A^\dagger(G_x)
        \qquad\text{for every }x.
        \label{eq:app_povm_simulation}
    \end{align}

    \item
    There exists a Hermiticity-preserving, trace-preserving statistical
    morphism $\mathcal L:\cL(A)\to\cL(E)$ of $\mathcal N_A$ such that
    \begin{align}
        \mathcal N_E
        =
        \mathcal L\circ\mathcal N_A.
        \label{eq:app_statistical_degradation}
    \end{align}
\end{enumerate}
Equivalently, Eq.~\eqref{eq:app_statistical_degradation} can be written
as
\begin{align}
    \rho_{EB}
    =
    (\mathcal L\otimes\id_B)(\rho_{AB}).
    \label{eq:app_state_statistical_degradation}
\end{align}
\end{theorem}

\begin{proof}
For an Eve POVM $\{F_x\}_x$, the subnormalized state prepared on $B$
for outcome $x$ is, by
Eq.~\eqref{eq:app_faithful_steering_identity},
\begin{align}
    \widetilde\eta_x
    &=    \Tr_E[(F_x\otimes\mathbb I_B)\rho_{EB}]
    =
    \sqrt{\rho_B}\,
    \mathcal N_E^\dagger(F_x)^T
    \sqrt{\rho_B}.
\end{align}
Similarly, an Alice POVM $\{G_x\}_x$ prepares
\begin{align}
    \widetilde\tau_x
    &=
    \Tr_A[(G_x\otimes\mathbb I_B)\rho_{AB}]
    =
    \sqrt{\rho_B}\,
    \mathcal N_A^\dagger(G_x)^T
    \sqrt{\rho_B}.
\end{align}
Since $\rho_B$ is full rank, multiplication by $\sqrt{\rho_B}$ is
invertible. Therefore
$\widetilde\eta_x=\widetilde\tau_x$ if and only if
\begin{align}
    \mathcal N_E^\dagger(F_x)
    =
    \mathcal N_A^\dagger(G_x).
\end{align}
Equality of the subnormalized states implies equality of both the
outcome probabilities and the corresponding normalized conditional
states. Hence every ensemble generated by an Eve POVM can be reproduced
by an Alice POVM if and only if the POVM-simulation condition
\eqref{eq:app_povm_simulation} holds. This proves the equivalence of
statements 1 and 2.

Assume statement 2. Choose a linearly independent informationally
complete POVM $\{F_j\}_{j=1}^{d_E^2}$ on $E$, and let
$\{G_j\}_{j=1}^{d_E^2}$ be a simulating POVM on $A$. Because the
$F_j$ form a Hermitian operator basis, the assignments
$\mathcal L^\dagger(F_j)=G_j$ define a unique Hermiticity-preserving
linear map $\mathcal L^\dagger:\cL(E)\to\cL(A)$. Moreover,
\begin{align}
    \mathcal L^\dagger(\mathbb I_E)
    =
    \sum_jG_j
    =
    \mathbb I_A,
\end{align}
so $\mathcal L$ is trace preserving. Equation
\eqref{eq:app_povm_simulation} on the basis $\{F_j\}$, followed by
linearity, gives
\begin{align}
    \mathcal N_E^\dagger
    =
    \mathcal N_A^\dagger\circ\mathcal L^\dagger,
\end{align}
which is Eq.~\eqref{eq:app_statistical_degradation}. It remains to
verify that $\mathcal L$ has the measurement-simulation property
required of a statistical morphism. Let $\{F_x\}_x$ be an arbitrary
POVM on $E$. By statement 2, there is a POVM $\{G_x\}_x$ on $A$ such
that
\begin{align}
    \mathcal N_E^\dagger(F_x)
    =
    \mathcal N_A^\dagger(G_x)
\end{align}
for every $x$. Using the factorization
$\mathcal N_E=\mathcal L\circ\mathcal N_A$, we obtain, for every input
state $\omega$,
\begin{align}
    \Tr[\mathcal L(\mathcal N_A(\omega))F_x]
    &=
    \Tr[\mathcal N_E(\omega)F_x]
    \nonumber\\
    &=
    \Tr[\omega\,\mathcal N_E^\dagger(F_x)]
    \nonumber\\
    &=
    \Tr[\omega\,\mathcal N_A^\dagger(G_x)]
    \nonumber\\
    &=
    \Tr[\mathcal N_A(\omega)G_x].
\end{align}
This is precisely the statistical-morphism condition
\eqref{eq:app_statistical_morphism}. Hence $\mathcal L$ is the required
statistical morphism.

Conversely, assume statement 3. For every Eve POVM, the defining
property of a statistical morphism gives an Alice POVM with identical
statistics on every output of $\mathcal N_A$. Using
$\mathcal N_E=\mathcal L\circ\mathcal N_A$ and taking adjoints gives
Eq.~\eqref{eq:app_povm_simulation}.

Finally, the map
$\mathcal Q\mapsto(\mathcal Q\otimes\id_B)(\psi_\rho)$ is injective
because $\rho_B$ is full rank. Hence the channel factorization
\eqref{eq:app_statistical_degradation} and the state relation
\eqref{eq:app_state_statistical_degradation} are equivalent.
\end{proof}

\subsection{Positive degradability}
\label{app:positive_degradability}

Positive trace-preserving degradability is sufficient for exact
ensemble simulation, but it is not equivalent to it in general.

\begin{proposition}
\label{prop:app_positive_sufficient}
If there is a positive trace-preserving map
$\mathcal P:\cL(A)\to\cL(E)$ satisfying
\begin{align}
    \rho_{EB}
    =
    (\mathcal P\otimes\id_B)(\rho_{AB}),
    \label{eq:app_ptp_state_relation}
\end{align}
then
$\Ens_{E\to B}(\rho_{EB})
\subseteq\Ens_{A\to B}(\rho_{AB})$.
\end{proposition}

\begin{proof}
The adjoint of a positive trace-preserving map is positive and unital.
Thus every Eve POVM $\{F_x\}_x$ is mapped to the Alice POVM
$\{\mathcal P^\dagger(F_x)\}_x$, and
Eq.~\eqref{eq:app_ptp_state_relation} shows that the two POVMs prepare
the same subnormalized states on $B$.
\end{proof}

A statistical morphism need only reproduce measurements on the range of $\mathcal N_A$ and need not admit a positive extension to all of $\cL(A)$. Explicit counterexamples are known~\cite{matsumoto2014example}.

\subsection{Rank-deficient Bob marginal}

If $\rho_B$ is not full rank, let $\Pi_B$ project onto
$\mathcal K_B=\supp\rho_B$. Positivity implies
\begin{align}
    \rho_{XB}
    =
    (\mathbb I_X\otimes\Pi_B)\rho_{XB}
    (\mathbb I_X\otimes\Pi_B),
    \qquad X=A,E.
\end{align}
Every remotely prepared state is therefore supported on
$\mathcal K_B$. Replacing $B$ by $\mathcal K_B$ leaves both ensemble
sets unchanged and makes $\rho_B$ full rank. The construction and
Theorem~\ref{thm:app_statistical_simulation} then apply on this support.

\section{H. PTP/CPTP separation}
\label{app:ptp_not_cptp}

This section establishes that universal PTP simulation can hold even
when neither direction is realizable by a physical CPTP degradable channel. Unlike an
arbitrary state-conversion example, the two states constructed below
are the actual complementary marginals of one pure state. The construction is adapted from the channel and explicit transpose
simulator of Zhu and Wang~\cite{zhu2026zero}.

Let $A\simeq\mathbb C^8$ and $B\simeq E\simeq\mathbb C^4$. In the
computational basis of the four-dimensional systems, define
\begin{align}
    D&=\operatorname{diag}(1,-4,4,-1),
    \nonumber\\
    A_1&=|0\rangle\!\langle1|-2|1\rangle\!\langle2|
          +|2\rangle\!\langle3|,
    \nonumber\\
    A_2&=|0\rangle\!\langle2|-|1\rangle\!\langle3|,
    & \nonumber \\
    A_3&=|0\rangle\!\langle3|.
\end{align}
Set
\begin{align}
    (L_0,\ldots,L_7)
    ={}&
    (\mathbb I,D,A_1,A_1^\dagger,A_2,A_2^\dagger,
      A_3,A_3^\dagger),
    \label{eq:app_ptp_operators}\\
    (w_0,\ldots,w_7)
    ={}&
    \frac1{56}(8,1,4,4,12,12,31,31),
    \label{eq:app_ptp_weights}\\
    K_e={}&\sqrt{w_e}\,L_e.
\end{align}
The identities
\begin{align}
    \sum_{e=0}^7K_e^\dagger K_e
    =
    \sum_{e=0}^7K_eK_e^\dagger
    =
    \mathbb I_4
\end{align}
show that
\begin{align}
    \mathcal N(Z)
    &=
    \sum_{e=0}^7K_e ZK_e^\dagger,
    \nonumber\\
    [\mathcal N^c(Z)]_{ef}
    &=
    \Tr(K_e ZK_f^\dagger)
    \label{eq:app_ptp_complementary_channels}
\end{align}
define complementary channels, with outputs $E$ and $A$, respectively,
and with $\mathcal N$ unital. A normalized common purification is
\begin{align}
    |\Psi\rangle_{ABE}
    =
    \frac12\sum_{j=0}^3\sum_{e=0}^7
    \sqrt{w_e}\,|e\rangle_A|j\rangle_B L_e|j\rangle_E.
    \label{eq:app_ptp_purification}
\end{align}
Its complementary marginals are
\begin{align}
    \rho_{AB}
    ={}&
    \frac14\sum_{e,f=0}^7\sqrt{w_ew_f}\,
    |e\rangle\!\langle f|_A
    \nonumber\\
    &\qquad\otimes
    (L_f^\dagger L_e)^{\mathsf T}_B,
    \label{eq:app_ptp_source}\\
    \rho_{EB}
    ={}&
    \sum_{e=0}^7w_e(L_e\otimes\mathbb I_B)
    \Phi_4
    (L_e^\dagger\otimes\mathbb I_B),
    \label{eq:app_ptp_complement}
\end{align}
where
$|\Phi_4\rangle_{EB}=\frac12\sum_{j=0}^3|j\rangle_E|j\rangle_B$
and $\Phi_4=|\Phi_4\rangle\!\langle\Phi_4|$. Both states have
$\rho_B=\mathbb I_4/4$.

\begin{proposition}
\label{prop:app_complementary_ptp_not_cptp}
The complementary states in
Eqs.~\eqref{eq:app_ptp_source} and \eqref{eq:app_ptp_complement}
satisfy the following properties:
\begin{enumerate}
    \item There is a PTP map $\mathcal Q_{A\to E}$ such that
    \begin{align}
        \rho_{EB}
        =
        (\mathcal Q_{A\to E}\otimes\id_B)(\rho_{AB}).
        \label{eq:app_ptp_state_conversion}
    \end{align}
    \item No CPTP map implements either this conversion or its
    reverse.
    \item For the purity functional
    $\mathcal P(\omega_B)=2-S(\omega_B)$,
    \begin{align}
        \Delta^{\mathcal P}(\rho_{AB})
        \geq
        0.0156484731609\ldots>0.
        \label{eq:app_ptp_secure_purity_bound}
    \end{align}
\end{enumerate}
Consequently,
\begin{align}
    \Ens_{E\to B}(\rho_{EB})
    \subsetneq
    \Ens_{A\to B}(\rho_{AB}).
    \label{eq:app_ptp_strict_ensemble_order}
\end{align}
This gives a strict simulation order that neither direction of CPTP
degradation detects.
\end{proposition}

\begin{proof}
We prove the three claims in turn.

\textit{Positive degradation.---}
Let $\mathsf T$ denote matrix transposition on $E$. The CPTP map
$\mathcal D_{A\to E}$ specified explicitly below obeys
\begin{align}
    \mathcal D\circ\mathcal N^c
    =
    \mathsf T\circ\mathcal N.
    \label{eq:app_ptp_transpose_simulation}
\end{align}
Therefore
\begin{align}
    \mathcal Q
    =
    \mathsf T\circ\mathcal D
    \quad\Longrightarrow\quad
    \mathcal Q\circ\mathcal N^c=\mathcal N.
    \label{eq:app_ptp_degrading_map}
\end{align}
Because transposition is positive and trace preserving, $\mathcal Q$
is PTP. Applying Eq.~\eqref{eq:app_ptp_degrading_map} to one half of
$\Phi_4$ proves Eq.~\eqref{eq:app_ptp_state_conversion}. Moreover,
$\mathcal Q^\dagger$ is positive and unital. Hence every Eve POVM
$\{F_x^E\}_x$ pulls back to the Alice POVM
$\{\mathcal Q^\dagger(F_x^E)\}_x$, which prepares the same
subnormalized states on $B$. This proves
\begin{align}
    \Ens_{E\to B}(\rho_{EB})
    \subseteq
    \Ens_{A\to B}(\rho_{AB}).
    \label{eq:app_ptp_ensemble_inclusion}
\end{align}

\textit{Absence of a CPTP degrading map.---}
Suppose that a CPTP map $\Gamma_{A\to E'}$, where $E'\simeq E$,
satisfied Eq.~\eqref{eq:app_ptp_state_conversion}. Applying $\Gamma$
to the $A$ part of $|\Psi\rangle_{ABE}$ would produce a positive state
$\Omega_{E'BE}$ whose $E'B$ and $EB$ marginals both equal
$\rho_{EB}$, after identifying $E'$ with $E$. We show that such an
extension cannot exist.

Let
\begin{align}
    \mathcal S
    =
    \operatorname{span}
    \bigl\{(L_e\otimes\mathbb I_B)|\Phi_4\rangle:
    0\leq e\leq7\bigr\}.
\end{align}
All the weights $w_e$ are positive, so $\mathcal S$ is precisely the
support of $\rho_{EB}$. It is invariant under swapping its two
four-dimensional systems because the list in
Eq.~\eqref{eq:app_ptp_operators} is closed under transposition. Any
positive extension having both required marginals must consequently be
supported, in the tensor order $BEE'$, on
\begin{align}
    \mathcal K
    =\bigl\{|t\rangle:;&
    (t_{ijk})_{i,j}\in\mathcal S_{BE}\ \text{for every }k,
    \nonumber\\[-1mm]
    &
    (t_{ijk})_{i,k}\in\mathcal S_{BE'}\ \text{for every }j
    \bigr\}.
    \label{eq:app_ptp_common_support}
\end{align}

We now establish the support obstruction
\begin{align}
    (\langle\Phi_4|_{BE}\otimes\mathbb I_{E'})
    \mathcal K
    =
    \{0\}.
    \label{eq:app_ptp_support_obstruction}
\end{align}
Write a vector in $\mathcal K$ as
$|t\rangle=\sum_{i,j,k}t_{ijk}|i\rangle_B|j\rangle_E|k\rangle_{E'}$.
For every fixed $k$, the $BE$ support condition implies
\begin{align}
    (t_{ijk})_{i,j}
    =
    \begin{pmatrix}
       a_k+b_k & c_k & d_k & e_k\\
       f_k & a_k-4b_k & -2c_k & -d_k\\
       g_k & -2f_k & a_k+4b_k & c_k\\
       h_k & -g_k & f_k & a_k-b_k
    \end{pmatrix}.
    \label{eq:app_ptp_slice_form}
\end{align}
The $BE'$ support condition requires the same matrix form for every
fixed $j$, now with row and column indices $i,k$. From the $j=1$
slice one obtains
\begin{align}
    f_1=a_0-4b_0,
    \qquad
    g_2=-a_0+4b_0.
\end{align}
The diagonal constraints in the $j=0$ slice then give
\begin{align}
    h_3&=-a_0-b_0,
    \nonumber\\
    0&=4(a_0+b_0)+f_1-g_2-4h_3=10a_0,
\end{align}
and hence $a_0=0$. The off-diagonal constraints from the $j=0$ and
$j=1$ slices give
\begin{align}
    f_3&=-a_2-b_2,
    &c_1&=-2f_3,
    &a_2-4b_2&=-2c_1,
\end{align}
so $a_2=0$. Simultaneously reversing all three basis indices preserves
the two support conditions and maps $a_k$ to $a_{3-k}$. Therefore
$a_1=a_3=0$ as well. It follows that
$\sum_i t_{iik}=4a_k=0$ for every $k$, proving
Eq.~\eqref{eq:app_ptp_support_obstruction}.

Every positive operator supported on $\mathcal K$ therefore has zero
overlap of its $EB$ marginal with $\Phi_4$. On the other hand,
\begin{align}
    \Tr(\Phi_4\rho_{EB})
    =
    \frac1{16}\sum_{e=0}^7w_e|\Tr L_e|^2
    =
    \frac17.
\end{align}
This contradiction excludes every CPTP degrading map $A\to E$.
It also shows that $\rho_{EB}$ has no symmetric two-extension and is
therefore entangled.

\textit{Absence of a CPTP antidegrading map.---}
The operators $L_e$ are Hilbert--Schmidt orthogonal. Hence
\begin{align}
    \rho_A
    &=
    \operatorname{diag}(\boldsymbol q),
    &
    \boldsymbol q
    &=
    \frac1{224}(32,34,24,24,24,24,31,31).
    \label{eq:app_ptp_A_spectrum}
\end{align}
Unitality of $\mathcal N$ gives $\rho_E=\mathbb I_4/4$. Since
$|\Psi\rangle_{ABE}$ is pure,
\begin{align}
    S(A)&=H(\boldsymbol q),
    &S(B)&=S(E)=S(AB)=2,
    \nonumber\\
    S(EB)&=H(\boldsymbol q),
\end{align}
where $H$ is the Shannon entropy in bits. Because
$\max_e q_e=17/112<1/4$,
\begin{align}
    H(\boldsymbol q)
    \geq
    -\log_2\!\left(\max_e q_e\right)
    =
    \log_2\!\left(\frac{112}{17}\right)>2.
\end{align}
If a CPTP map $E\to A$ converted $\rho_{EB}$ into $\rho_{AB}$, data
processing of the mutual information would require
\begin{align}
    I(B{:}A)_\rho
    =H(\boldsymbol q)
    \leq
    I(B{:}E)_\rho
    =4-H(\boldsymbol q),
\end{align}
which is impossible. Thus the state is neither CPTP degradable nor
CPTP antidegradable. In addition,
$S(B|A)_\rho=2-H(\boldsymbol q)<0$, so $\rho_{AB}$ is entangled.

\textit{Strict exclusive-purity advantage.---}
Equation~\eqref{eq:app_purity_delta} gives
\begin{align}
    \Delta^{\mathcal P}(\rho_{AB})
    =
    E_F(\rho_{AB})-E_F(\rho_{EB}).
\end{align}
Concavity of conditional entropy implies
\begin{align}
    E_F(\rho_{AB})
    \geq
    -S(B|A)_\rho
    =
    H(\boldsymbol q)-2
    \nonumber\\
    =
    0.9846303331481005\ldots.
    \label{eq:app_ptp_source_EF_lower}
\end{align}
For the complementary state, use the pure-state ensemble displayed in
Eq.~\eqref{eq:app_ptp_complement}. Its normalized vectors occur with
probabilities
\begin{align}
    q_e
    =
    \frac{w_e}{4}\Tr(L_e^\dagger L_e),
\end{align}
and their entanglement entropies, in order $e=0,\ldots,7$, are
\begin{align}
    \bigl(&2,\ 1+h_2(1/17),\ h_2(1/3)+1/3,
    \nonumber\\
          &h_2(1/3)+1/3,\ 1,\ 1,\ 0,\ 0\bigr),
\end{align}
where $h_2(x)=-x\log_2x-(1-x)\log_2(1-x)$. Consequently,
\begin{align}
    E_F(\rho_{EB})
    \leq U
    :={}&
    \frac27
    +\frac{17}{112}\bigl[1+h_2(1/17)\bigr]
    \nonumber\\
    &+\frac3{14}\bigl[h_2(1/3)+1/3\bigr]
    +\frac3{14}
    \nonumber\\
    ={}&0.9689818599871742\ldots.
    \label{eq:app_ptp_complement_EF_upper}
\end{align}
Combining Eqs.~\eqref{eq:app_ptp_source_EF_lower} and
\eqref{eq:app_ptp_complement_EF_upper} yields
\begin{align}
    \Delta^{\mathcal P}(\rho_{AB})
    \geq{}&
    H(\boldsymbol q)-2-U
    \nonumber\\
    ={}&
    \frac{1}{112}
    \log_2\!\left(
    \frac{2^{94}7^{112}}
         {3^{72}17^{34}31^{31}}
    \right)
    \nonumber\\
    ={}&0.0156484731609262\ldots>0,
\end{align}
where positivity may also be checked directly from
$2^{94}7^{112}>3^{72}17^{34}31^{31}$. This proves
Eq.~\eqref{eq:app_ptp_secure_purity_bound}. It also makes the ensemble
inclusion strict, since equality of the two ensemble sets would force
equal optimized scores for every task.
\end{proof}

\subsection{Explicit certificate for the transpose simulator}

For completeness, we specify the CPTP map $\mathcal D$ used in
Eq.~\eqref{eq:app_ptp_transpose_simulation}. In the basis
$|e,c\rangle$ of $A\otimes E$, define
\begin{align}
    |v_1\rangle
    &=
    \frac12|0,0\rangle-\frac18|1,0\rangle+|3,1\rangle,
    \nonumber\\
    |v_2\rangle
    &=
    \frac12|0,0\rangle+\frac18|1,0\rangle+|5,2\rangle,
    \nonumber\\
    |v_3\rangle&=|7,3\rangle,
    \nonumber\\
    |v\rangle
    &=
    2|0,2\rangle+|1,2\rangle-2|2,1\rangle
    \nonumber\\
    &\qquad+|3,3\rangle+3|4,0\rangle,
    \nonumber\\
    |a\rangle
    &=
    |3,0\rangle-2|5,1\rangle+|7,2\rangle,
    \nonumber\\
    |b\rangle&=-|5,0\rangle+|7,1\rangle,
    &|z\rangle&=|7,0\rangle.
    \label{eq:app_ptp_simulator_vectors}
\end{align}
Let $V=(|v_1\rangle,|v_2\rangle,|v_3\rangle)$ and set
\begin{align}
    M&=
    \begin{pmatrix}
        16&8&29\\
        8&64&77\\
        29&77&251
    \end{pmatrix},
    \nonumber\\
    Q&=
    VMV^\dagger
    +4\bigl(vv^\dagger+aa^\dagger+bb^\dagger\bigr)
    +51zz^\dagger.
    \label{eq:app_ptp_Q}
\end{align}
The leading principal minors of $M$ are $16$, $960$, and $128000$;
hence $M\succ0$ and $Q\succeq0$. Define the orthogonal operator
$\mathsf F$ by
\begin{align}
    \mathsf F|e,c\rangle
    &=
    s_e|\pi(e),3-c\rangle,
    \nonumber\\
    \pi&=(0,1,3,2,5,4,7,6),
    \nonumber\\
    s&=(1,-1,1,1,-1,-1,1,1),
\end{align}
where the two lists give $\pi(e)$ and $s_e$ for $e=0,\ldots,7$.
Finally, let
\begin{align}
    W&=\operatorname{diag}(\sqrt{w_0},\ldots,\sqrt{w_7}),
    \nonumber\\
    H&=\frac{Q+\mathsf FQ\mathsf F^\dagger}{560},
    \nonumber\\
    J_{\mathcal D}
    &=
    (W^{-1}\otimes\mathbb I_4)
    H
    (W^{-1}\otimes\mathbb I_4).
    \label{eq:app_ptp_simulator_choi}
\end{align}
We use the unnormalized, input-first Choi convention
\begin{align}
    J_{\mathcal D}
    =
    \sum_{e,f=0}^7|e\rangle\!\langle f|
    \otimes\mathcal D(|e\rangle\!\langle f|).
\end{align}
If $H_{ef}$ denotes the $(e,f)$ block of $H$, then
\begin{align}
    \mathcal D(Z)
    =
    \sum_{e,f=0}^7
    \frac{Z_{ef}}{\sqrt{w_ew_f}}H_{ef}.
\end{align}
The positive factorization of $H$ proves $J_{\mathcal D}\succeq0$.
The remaining claims follow from the finite identities
\begin{align}
    \Tr_E H
    &=
    \operatorname{diag}(w_0,\ldots,w_7),
    \label{eq:app_ptp_simulator_TP}\\
    \sum_{e,f=0}^7
    (L_f^\dagger L_e)_{ji}H_{ef}
    &=
    \mathcal N(|i\rangle\!\langle j|)^{\mathsf T},
    \quad 0\leq i,j\leq3.
    \label{eq:app_ptp_simulator_composition}
\end{align}
Equation~\eqref{eq:app_ptp_simulator_TP} proves that $\mathcal D$ is
trace preserving, while
Eq.~\eqref{eq:app_ptp_simulator_composition} proves
Eq.~\eqref{eq:app_ptp_transpose_simulation}. All entries in these
identities are rational, so the certificate can be checked exactly.

\section{I. Rank-two order and amplitude damping}\label{app:amplitude_damping}

For this family the two exact ensemble orders meet at the
self-complementary point, and all three plotted task advantages can be
evaluated analytically. Consider
\begin{align}
    |\Psi_\gamma\rangle_{ABE}
    =
    \frac{
        |000\rangle
        +\sqrt{1-\gamma}|110\rangle
        +\sqrt{\gamma}|011\rangle
    }{\sqrt2},
    \qquad 0\leq\gamma\leq1.
    \label{eq:amplitude-purification}
\end{align}
With $|\chi_q\rangle=|00\rangle+\sqrt q\,|11\rangle$, its
complementary marginals are
\begin{align}
    \rho_{AB}
    &=
    \frac12|\chi_{1-\gamma}\rangle
    \!\langle\chi_{1-\gamma}|
    +\frac{\gamma}{2}|01\rangle\!\langle01|,
    \nonumber\\
    \rho_{EB}
    &=
    \frac12|\chi_{\gamma}\rangle
    \!\langle\chi_{\gamma}|
    +\frac{1-\gamma}{2}|01\rangle\!\langle01|.
    \label{eq:app_amplitude_marginals}
\end{align}
They have rank two for $0<\gamma<1$ (and rank at most two at the
endpoints) and concurrences
\begin{align}
    C(\rho_{AB})=\sqrt{1-\gamma},
    \qquad
    C(\rho_{EB})=\sqrt{\gamma}.
\end{align}
Writing
$h_2(x)=-x\log_2x-(1-x)\log_2(1-x)$, Wootters' formula gives
\begin{align}
    E_F(\rho_{AB})
    &=h_2\!\left(\frac{1+\sqrt\gamma}{2}\right),
    \nonumber\\
    E_F(\rho_{EB})
    &=h_2\!\left(\frac{1+\sqrt{1-\gamma}}{2}\right).
    \label{eq:app_amplitude_eof}
\end{align}
For $H_B=\varepsilon|1\rangle\!\langle1|$, the exclusive advantages are
\begin{align}
    \Delta^{\mathcal P}
    &=
    E_F(\rho_{AB})-E_F(\rho_{EB}),
    \nonumber\\
    \Delta^{\mathcal W_\beta}
    &=k_BT\ln2\,\Delta^{\mathcal P},
    \nonumber\\
    \Delta^{\mathcal W_{\rm erg}}
    &=
    \frac{\varepsilon}{2}
    \left(\sqrt{1-\gamma}-\sqrt\gamma\right),
    \label{eq:amplitude-advantages}
\end{align}
where $\Delta^{\mathcal P}$ is reported in bits in this display. For the ergotropy, the energy term is affine and cancels, and the
remaining score is $\frac{\varepsilon}{2}\max\sum_kp_k|\boldsymbol r_k|
=\frac{\varepsilon}{2}\max\sum_kp_k\sqrt{1-C(\phi_k)^2}$. Since
$1-\sqrt{1-C^2}$ is convex and increasing in $C$, Wootters' optimal
decomposition gives the maximum $\frac{\varepsilon}{2}\sqrt{1-C(\sigma)^2}$,
i.e., $\frac{\varepsilon}{2}\sqrt{1-\gamma}$ for Alice and
$\frac{\varepsilon}{2}\sqrt{\gamma}$ for Eve.

The coherence curve in the End Matter Sec.~\ref{sec:end_task_independent} is exact as well. For
$q\in[0,1]$, define
\begin{align}
    \sigma_{XB}(q)
    =
    \frac12|\chi_q\rangle\!\langle\chi_q|
    +\frac{1-q}{2}|01\rangle\!\langle01|.
\end{align}
Writing $S_2$ for the entropy in bits, every pure-state decomposition obeys
\begin{align}
    \sum_kp_k C_{\rm r}(\phi_B^k)
    \leq
    1-\sum_kp_kS_2(\phi_B^k)
    \leq
    1-E_F(\sigma_{XB}(q)).
\end{align}
The upper bound is attained by the two-state decomposition
\begin{align}
    |\varphi_\pm(q)\rangle
    =
    \frac{
        |00\rangle+\sqrt q\,|11\rangle
        \pm\sqrt{1-q}\,|01\rangle
    }{\sqrt2},
\end{align}
because each Bob marginal has a uniform incoherent-basis diagonal and
each state has concurrence $\sqrt q$. Therefore
\begin{align}
    \widehat C_{\rm r}(\sigma_{XB}(q))
    =1-E_F(\sigma_{XB}(q)),
\end{align}
and substitution of $q=1-\gamma$ and $q=\gamma$ gives
\begin{align}
    \Delta^{C_{\rm r}}
    =E_F(\rho_{AB})-E_F(\rho_{EB}).
    \label{eq:app_amplitude_coherence}
\end{align}
Thus all three signed advantages are positive for
$0\leq\gamma<1/2$, vanish at $\gamma=1/2$, and are negative for
$1/2<\gamma\leq1$.

Finally, the channel to $A$ is the amplitude-damping channel
$\mathcal D_\gamma$, while its complement is unitarily equivalent to
$\mathcal D_{1-\gamma}$. Since
$\mathcal D_a\circ\mathcal D_b=\mathcal D_{a+b-ab}$,
\begin{align}
    \mathcal D_{1-\gamma}
    &=
    \mathcal D_{\frac{1-2\gamma}{1-\gamma}}
    \circ\mathcal D_\gamma,
    &&0\leq\gamma\leq\frac12,
    \nonumber\\
    \mathcal D_\gamma
    &=
    \mathcal D_{\frac{2\gamma-1}{\gamma}}
    \circ\mathcal D_{1-\gamma},
    &&\frac12\leq\gamma\leq1.
    \label{eq:app_amplitude_degradation}
\end{align}
Hence
$\Ens_{E\to B}\subseteq\Ens_{A\to B}$ for
$0\leq\gamma\leq1/2$, while the reverse inclusion holds for
$1/2\leq\gamma\leq1$. At $\gamma=1/2$ both channels are
self-complementary and the ensemble sets coincide.


\begin{thebibliography}{44}%
\makeatletter
\providecommand \@ifxundefined [1]{%
 \@ifx{#1\undefined}
}%
\providecommand \@ifnum [1]{%
 \ifnum #1\expandafter \@firstoftwo
 \else \expandafter \@secondoftwo
 \fi
}%
\providecommand \@ifx [1]{%
 \ifx #1\expandafter \@firstoftwo
 \else \expandafter \@secondoftwo
 \fi
}%
\providecommand \natexlab [1]{#1}%
\providecommand \enquote  [1]{``#1''}%
\providecommand \bibnamefont  [1]{#1}%
\providecommand \bibfnamefont [1]{#1}%
\providecommand \citenamefont [1]{#1}%
\providecommand \href@noop [0]{\@secondoftwo}%
\providecommand \href [0]{\begingroup \@sanitize@url \@href}%
\providecommand \@href[1]{\@@startlink{#1}\@@href}%
\providecommand \@@href[1]{\endgroup#1\@@endlink}%
\providecommand \@sanitize@url [0]{\catcode `\\12\catcode `\$12\catcode `\&12\catcode `\#12\catcode `\^12\catcode `\_12\catcode `\%12\relax}%
\providecommand \@@startlink[1]{}%
\providecommand \@@endlink[0]{}%
\providecommand \url  [0]{\begingroup\@sanitize@url \@url }%
\providecommand \@url [1]{\endgroup\@href {#1}{\urlprefix }}%
\providecommand \urlprefix  [0]{URL }%
\providecommand \Eprint [0]{\href }%
\providecommand \doibase [0]{https://doi.org/}%
\providecommand \selectlanguage [0]{\@gobble}%
\providecommand \bibinfo  [0]{\@secondoftwo}%
\providecommand \bibfield  [0]{\@secondoftwo}%
\providecommand \translation [1]{[#1]}%
\providecommand \BibitemOpen [0]{}%
\providecommand \bibitemStop [0]{}%
\providecommand \bibitemNoStop [0]{.\EOS\space}%
\providecommand \EOS [0]{\spacefactor3000\relax}%
\providecommand \BibitemShut  [1]{\csname bibitem#1\endcsname}%
\let\auto@bib@innerbib\@empty
\bibitem [{\citenamefont {Preskill}(2018)}]{preskill2018quantum}%
  \BibitemOpen
  \bibfield  {author} {\bibinfo {author} {\bibfnamefont {J.}~\bibnamefont {Preskill}},\ }\bibfield  {title} {\bibinfo {title} {Quantum {C}omputing in the {NISQ} era and beyond},\ }\href {https://doi.org/10.22331/q-2018-08-06-79} {\bibfield  {journal} {\bibinfo  {journal} {{Quantum}}\ }\textbf {\bibinfo {volume} {2}},\ \bibinfo {pages} {79} (\bibinfo {year} {2018})}\BibitemShut {NoStop}%
\bibitem [{\citenamefont {Devitt}(2016)}]{devitt2016performing}%
  \BibitemOpen
  \bibfield  {author} {\bibinfo {author} {\bibfnamefont {S.~J.}\ \bibnamefont {Devitt}},\ }\bibfield  {title} {\bibinfo {title} {Performing quantum computing experiments in the cloud},\ }\href {https://doi.org/10.1103/PhysRevA.94.032329} {\bibfield  {journal} {\bibinfo  {journal} {Phys. Rev. A}\ }\textbf {\bibinfo {volume} {94}},\ \bibinfo {pages} {032329} (\bibinfo {year} {2016})}\BibitemShut {NoStop}%
\bibitem [{\citenamefont {Wehner}\ \emph {et~al.}(2018)\citenamefont {Wehner}, \citenamefont {Elkouss},\ and\ \citenamefont {Hanson}}]{wehner2018quantum}%
  \BibitemOpen
  \bibfield  {author} {\bibinfo {author} {\bibfnamefont {S.}~\bibnamefont {Wehner}}, \bibinfo {author} {\bibfnamefont {D.}~\bibnamefont {Elkouss}},\ and\ \bibinfo {author} {\bibfnamefont {R.}~\bibnamefont {Hanson}},\ }\bibfield  {title} {\bibinfo {title} {Quantum internet: A vision for the road ahead},\ }\href {https://doi.org/10.1126/science.aam9288} {\bibfield  {journal} {\bibinfo  {journal} {Science}\ }\textbf {\bibinfo {volume} {362}},\ \bibinfo {pages} {eaam9288} (\bibinfo {year} {2018})}\BibitemShut {NoStop}%
\bibitem [{\citenamefont {Komar}\ \emph {et~al.}(2014)\citenamefont {Komar}, \citenamefont {Kessler}, \citenamefont {Bishof}, \citenamefont {Jiang}, \citenamefont {S{\o}rensen}, \citenamefont {Ye},\ and\ \citenamefont {Lukin}}]{komar2014quantum}%
  \BibitemOpen
  \bibfield  {author} {\bibinfo {author} {\bibfnamefont {P.}~\bibnamefont {Komar}}, \bibinfo {author} {\bibfnamefont {E.~M.}\ \bibnamefont {Kessler}}, \bibinfo {author} {\bibfnamefont {M.}~\bibnamefont {Bishof}}, \bibinfo {author} {\bibfnamefont {L.}~\bibnamefont {Jiang}}, \bibinfo {author} {\bibfnamefont {A.~S.}\ \bibnamefont {S{\o}rensen}}, \bibinfo {author} {\bibfnamefont {J.}~\bibnamefont {Ye}},\ and\ \bibinfo {author} {\bibfnamefont {M.~D.}\ \bibnamefont {Lukin}},\ }\bibfield  {title} {\bibinfo {title} {A quantum network of clocks},\ }\href {https://doi.org/10.1038/nphys3000} {\bibfield  {journal} {\bibinfo  {journal} {Nat. Phys.}\ }\textbf {\bibinfo {volume} {10}},\ \bibinfo {pages} {582} (\bibinfo {year} {2014})}\BibitemShut {NoStop}%
\bibitem [{\citenamefont {Gisin}\ \emph {et~al.}(2002)\citenamefont {Gisin}, \citenamefont {Ribordy}, \citenamefont {Tittel},\ and\ \citenamefont {Zbinden}}]{gisin2002quantum}%
  \BibitemOpen
  \bibfield  {author} {\bibinfo {author} {\bibfnamefont {N.}~\bibnamefont {Gisin}}, \bibinfo {author} {\bibfnamefont {G.}~\bibnamefont {Ribordy}}, \bibinfo {author} {\bibfnamefont {W.}~\bibnamefont {Tittel}},\ and\ \bibinfo {author} {\bibfnamefont {H.}~\bibnamefont {Zbinden}},\ }\bibfield  {title} {\bibinfo {title} {Quantum cryptography},\ }\href {https://doi.org/10.1103/RevModPhys.74.145} {\bibfield  {journal} {\bibinfo  {journal} {Rev. Mod. Phys.}\ }\textbf {\bibinfo {volume} {74}},\ \bibinfo {pages} {145} (\bibinfo {year} {2002})}\BibitemShut {NoStop}%
\bibitem [{\citenamefont {Scarani}\ \emph {et~al.}(2009)\citenamefont {Scarani}, \citenamefont {Bechmann-Pasquinucci}, \citenamefont {Cerf}, \citenamefont {Du\ifmmode~\check{s}\else \v{s}\fi{}ek}, \citenamefont {L\"utkenhaus},\ and\ \citenamefont {Peev}}]{scarani2009security}%
  \BibitemOpen
  \bibfield  {author} {\bibinfo {author} {\bibfnamefont {V.}~\bibnamefont {Scarani}}, \bibinfo {author} {\bibfnamefont {H.}~\bibnamefont {Bechmann-Pasquinucci}}, \bibinfo {author} {\bibfnamefont {N.~J.}\ \bibnamefont {Cerf}}, \bibinfo {author} {\bibfnamefont {M.}~\bibnamefont {Du\ifmmode~\check{s}\else \v{s}\fi{}ek}}, \bibinfo {author} {\bibfnamefont {N.}~\bibnamefont {L\"utkenhaus}},\ and\ \bibinfo {author} {\bibfnamefont {M.}~\bibnamefont {Peev}},\ }\bibfield  {title} {\bibinfo {title} {The security of practical quantum key distribution},\ }\href {https://doi.org/10.1103/RevModPhys.81.1301} {\bibfield  {journal} {\bibinfo  {journal} {Rev. Mod. Phys.}\ }\textbf {\bibinfo {volume} {81}},\ \bibinfo {pages} {1301} (\bibinfo {year} {2009})}\BibitemShut {NoStop}%
\bibitem [{\citenamefont {Pironio}\ \emph {et~al.}(2010)\citenamefont {Pironio}, \citenamefont {Ac{\'i}n}, \citenamefont {Massar}, \citenamefont {Boyer de~la Giroday}, \citenamefont {Matsukevich}, \citenamefont {Maunz}, \citenamefont {Olmschenk}, \citenamefont {Hayes}, \citenamefont {Luo}, \citenamefont {Manning},\ and\ \citenamefont {Monroe}}]{pironio2010randomness}%
  \BibitemOpen
  \bibfield  {author} {\bibinfo {author} {\bibfnamefont {S.}~\bibnamefont {Pironio}}, \bibinfo {author} {\bibfnamefont {A.}~\bibnamefont {Ac{\'i}n}}, \bibinfo {author} {\bibfnamefont {S.}~\bibnamefont {Massar}}, \bibinfo {author} {\bibfnamefont {A.}~\bibnamefont {Boyer de~la Giroday}}, \bibinfo {author} {\bibfnamefont {D.~N.}\ \bibnamefont {Matsukevich}}, \bibinfo {author} {\bibfnamefont {P.}~\bibnamefont {Maunz}}, \bibinfo {author} {\bibfnamefont {S.}~\bibnamefont {Olmschenk}}, \bibinfo {author} {\bibfnamefont {D.}~\bibnamefont {Hayes}}, \bibinfo {author} {\bibfnamefont {L.}~\bibnamefont {Luo}}, \bibinfo {author} {\bibfnamefont {T.~A.}\ \bibnamefont {Manning}},\ and\ \bibinfo {author} {\bibfnamefont {C.}~\bibnamefont {Monroe}},\ }\bibfield  {title} {\bibinfo {title} {Random numbers certified by {Bell}'s theorem},\ }\href {https://doi.org/10.1038/nature09008} {\bibfield  {journal} {\bibinfo  {journal} {Nature}\ }\textbf {\bibinfo {volume} {464}},\ \bibinfo {pages} {1021} (\bibinfo {year} {2010})}\BibitemShut
  {NoStop}%
\bibitem [{\citenamefont {Ac{\'i}n}\ and\ \citenamefont {Masanes}(2016)}]{acin2016}%
  \BibitemOpen
  \bibfield  {author} {\bibinfo {author} {\bibfnamefont {A.}~\bibnamefont {Ac{\'i}n}}\ and\ \bibinfo {author} {\bibfnamefont {L.}~\bibnamefont {Masanes}},\ }\bibfield  {title} {\bibinfo {title} {Certified randomness in quantum physics},\ }\href {https://doi.org/10.1038/nature20119} {\bibfield  {journal} {\bibinfo  {journal} {Nature}\ }\textbf {\bibinfo {volume} {540}},\ \bibinfo {pages} {213} (\bibinfo {year} {2016})}\BibitemShut {NoStop}%
\bibitem [{\citenamefont {Broadbent}\ \emph {et~al.}(2009)\citenamefont {Broadbent}, \citenamefont {Fitzsimons},\ and\ \citenamefont {Kashefi}}]{broadbent2009universal}%
  \BibitemOpen
  \bibfield  {author} {\bibinfo {author} {\bibfnamefont {A.}~\bibnamefont {Broadbent}}, \bibinfo {author} {\bibfnamefont {J.}~\bibnamefont {Fitzsimons}},\ and\ \bibinfo {author} {\bibfnamefont {E.}~\bibnamefont {Kashefi}},\ }\bibfield  {title} {\bibinfo {title} {Universal blind quantum computation},\ }in\ \href {https://doi.org/10.1109/FOCS.2009.36} {\emph {\bibinfo {booktitle} {2009 50th annual IEEE symposium on foundations of computer science}}}\ (\bibinfo {organization} {IEEE},\ \bibinfo {year} {2009})\ pp.\ \bibinfo {pages} {517--526}\BibitemShut {NoStop}%
\bibitem [{\citenamefont {Fitzsimons}(2017)}]{fitzsimons2017private}%
  \BibitemOpen
  \bibfield  {author} {\bibinfo {author} {\bibfnamefont {J.~F.}\ \bibnamefont {Fitzsimons}},\ }\bibfield  {title} {\bibinfo {title} {Private quantum computation: an introduction to blind quantum computing and related protocols},\ }\href {https://doi.org/10.1038/s41534-017-0025-3} {\bibfield  {journal} {\bibinfo  {journal} {npj Quantum Inf.}\ }\textbf {\bibinfo {volume} {3}},\ \bibinfo {pages} {23} (\bibinfo {year} {2017})}\BibitemShut {NoStop}%
\bibitem [{\citenamefont {Coecke}\ \emph {et~al.}(2016)\citenamefont {Coecke}, \citenamefont {Fritz},\ and\ \citenamefont {Spekkens}}]{coecke2016mathematical}%
  \BibitemOpen
  \bibfield  {author} {\bibinfo {author} {\bibfnamefont {B.}~\bibnamefont {Coecke}}, \bibinfo {author} {\bibfnamefont {T.}~\bibnamefont {Fritz}},\ and\ \bibinfo {author} {\bibfnamefont {R.~W.}\ \bibnamefont {Spekkens}},\ }\bibfield  {title} {\bibinfo {title} {A mathematical theory of resources},\ }\href {https://doi.org/10.1016/j.ic.2016.02.008} {\bibfield  {journal} {\bibinfo  {journal} {Inf. Comput.}\ }\textbf {\bibinfo {volume} {250}},\ \bibinfo {pages} {59} (\bibinfo {year} {2016})}\BibitemShut {NoStop}%
\bibitem [{\citenamefont {Chitambar}\ and\ \citenamefont {Gour}(2019)}]{chitambar2019quantum}%
  \BibitemOpen
  \bibfield  {author} {\bibinfo {author} {\bibfnamefont {E.}~\bibnamefont {Chitambar}}\ and\ \bibinfo {author} {\bibfnamefont {G.}~\bibnamefont {Gour}},\ }\bibfield  {title} {\bibinfo {title} {Quantum resource theories},\ }\href {https://doi.org/10.1103/RevModPhys.91.025001} {\bibfield  {journal} {\bibinfo  {journal} {Rev. Mod. Phys.}\ }\textbf {\bibinfo {volume} {91}},\ \bibinfo {pages} {025001} (\bibinfo {year} {2019})}\BibitemShut {NoStop}%
\bibitem [{\citenamefont {Chitambar}\ \emph {et~al.}(2016)\citenamefont {Chitambar}, \citenamefont {Streltsov}, \citenamefont {Rana}, \citenamefont {Bera}, \citenamefont {Adesso},\ and\ \citenamefont {Lewenstein}}]{chitambar2016assisted}%
  \BibitemOpen
  \bibfield  {author} {\bibinfo {author} {\bibfnamefont {E.}~\bibnamefont {Chitambar}}, \bibinfo {author} {\bibfnamefont {A.}~\bibnamefont {Streltsov}}, \bibinfo {author} {\bibfnamefont {S.}~\bibnamefont {Rana}}, \bibinfo {author} {\bibfnamefont {M.~N.}\ \bibnamefont {Bera}}, \bibinfo {author} {\bibfnamefont {G.}~\bibnamefont {Adesso}},\ and\ \bibinfo {author} {\bibfnamefont {M.}~\bibnamefont {Lewenstein}},\ }\bibfield  {title} {\bibinfo {title} {Assisted distillation of quantum coherence},\ }\href {https://doi.org/10.1103/PhysRevLett.116.070402} {\bibfield  {journal} {\bibinfo  {journal} {Phys. Rev. Lett.}\ }\textbf {\bibinfo {volume} {116}},\ \bibinfo {pages} {070402} (\bibinfo {year} {2016})}\BibitemShut {NoStop}%
\bibitem [{\citenamefont {Morris}\ \emph {et~al.}(2019)\citenamefont {Morris}, \citenamefont {Lami},\ and\ \citenamefont {Adesso}}]{morris2019assisted}%
  \BibitemOpen
  \bibfield  {author} {\bibinfo {author} {\bibfnamefont {B.}~\bibnamefont {Morris}}, \bibinfo {author} {\bibfnamefont {L.}~\bibnamefont {Lami}},\ and\ \bibinfo {author} {\bibfnamefont {G.}~\bibnamefont {Adesso}},\ }\bibfield  {title} {\bibinfo {title} {Assisted work distillation},\ }\href {https://doi.org/10.1103/PhysRevLett.122.130601} {\bibfield  {journal} {\bibinfo  {journal} {Phys. Rev. Lett.}\ }\textbf {\bibinfo {volume} {122}},\ \bibinfo {pages} {130601} (\bibinfo {year} {2019})}\BibitemShut {NoStop}%
\bibitem [{\citenamefont {Francica}\ \emph {et~al.}(2017)\citenamefont {Francica}, \citenamefont {Goold}, \citenamefont {Plastina},\ and\ \citenamefont {Paternostro}}]{francica2017daemonic}%
  \BibitemOpen
  \bibfield  {author} {\bibinfo {author} {\bibfnamefont {G.}~\bibnamefont {Francica}}, \bibinfo {author} {\bibfnamefont {J.}~\bibnamefont {Goold}}, \bibinfo {author} {\bibfnamefont {F.}~\bibnamefont {Plastina}},\ and\ \bibinfo {author} {\bibfnamefont {M.}~\bibnamefont {Paternostro}},\ }\bibfield  {title} {\bibinfo {title} {Daemonic ergotropy: Enhanced work extraction from quantum correlations},\ }\href {https://doi.org/10.1038/s41534-017-0012-8} {\bibfield  {journal} {\bibinfo  {journal} {npj Quantum Inf.}\ }\textbf {\bibinfo {volume} {3}},\ \bibinfo {pages} {12} (\bibinfo {year} {2017})}\BibitemShut {NoStop}%
\bibitem [{\citenamefont {Alimuddin}\ \emph {et~al.}(2025)\citenamefont {Alimuddin}, \citenamefont {Shettell}, \citenamefont {Yehia}, \citenamefont {Acín},\ and\ \citenamefont {Centrone}}]{alimuddin2025}%
  \BibitemOpen
  \bibfield  {author} {\bibinfo {author} {\bibfnamefont {M.}~\bibnamefont {Alimuddin}}, \bibinfo {author} {\bibfnamefont {N.}~\bibnamefont {Shettell}}, \bibinfo {author} {\bibfnamefont {R.}~\bibnamefont {Yehia}}, \bibinfo {author} {\bibfnamefont {A.}~\bibnamefont {Acín}},\ and\ \bibinfo {author} {\bibfnamefont {F.}~\bibnamefont {Centrone}},\ }\href {https://arxiv.org/abs/2512.05761} {\bibinfo {title} {Exclusive control of quantum memory erasure}} (\bibinfo {year} {2025}),\ \Eprint {https://arxiv.org/abs/2512.05761} {arXiv:2512.05761 [quant-ph]} \BibitemShut {NoStop}%
\bibitem [{\citenamefont {Buscemi}(2012)}]{buscemi2012comparison}%
  \BibitemOpen
  \bibfield  {author} {\bibinfo {author} {\bibfnamefont {F.}~\bibnamefont {Buscemi}},\ }\bibfield  {title} {\bibinfo {title} {Comparison of quantum statistical models: Equivalent conditions for sufficiency},\ }\href {https://doi.org/10.1007/s00220-012-1421-3} {\bibfield  {journal} {\bibinfo  {journal} {Communications in Mathematical Physics}\ }\textbf {\bibinfo {volume} {310}},\ \bibinfo {pages} {625} (\bibinfo {year} {2012})}\BibitemShut {NoStop}%
\bibitem [{\citenamefont {Buscemi}(2016)}]{buscemi2016degradable}%
  \BibitemOpen
  \bibfield  {author} {\bibinfo {author} {\bibfnamefont {F.}~\bibnamefont {Buscemi}},\ }\bibfield  {title} {\bibinfo {title} {Degradable channels, less noisy channels, and quantum statistical morphisms: An equivalence relation},\ }\href {https://doi.org/10.1134/S0032946016030017} {\bibfield  {journal} {\bibinfo  {journal} {Probl. Inf. Transm.}\ }\textbf {\bibinfo {volume} {52}},\ \bibinfo {pages} {201} (\bibinfo {year} {2016})}\BibitemShut {NoStop}%
\bibitem [{\citenamefont {Hughston}\ \emph {et~al.}(1993)\citenamefont {Hughston}, \citenamefont {Jozsa},\ and\ \citenamefont {Wootters}}]{hughston1993complete}%
  \BibitemOpen
  \bibfield  {author} {\bibinfo {author} {\bibfnamefont {L.~P.}\ \bibnamefont {Hughston}}, \bibinfo {author} {\bibfnamefont {R.}~\bibnamefont {Jozsa}},\ and\ \bibinfo {author} {\bibfnamefont {W.~K.}\ \bibnamefont {Wootters}},\ }\bibfield  {title} {\bibinfo {title} {A complete classification of quantum ensembles having a given density matrix},\ }\href {https://doi.org/10.1016/0375-9601(93)90880-9} {\bibfield  {journal} {\bibinfo  {journal} {Physics Letters A}\ }\textbf {\bibinfo {volume} {183}},\ \bibinfo {pages} {14} (\bibinfo {year} {1993})}\BibitemShut {NoStop}%
\bibitem [{\citenamefont {Wiseman}\ \emph {et~al.}(2007)\citenamefont {Wiseman}, \citenamefont {Jones},\ and\ \citenamefont {Doherty}}]{wiseman2007steering}%
  \BibitemOpen
  \bibfield  {author} {\bibinfo {author} {\bibfnamefont {H.~M.}\ \bibnamefont {Wiseman}}, \bibinfo {author} {\bibfnamefont {S.~J.}\ \bibnamefont {Jones}},\ and\ \bibinfo {author} {\bibfnamefont {A.~C.}\ \bibnamefont {Doherty}},\ }\bibfield  {title} {\bibinfo {title} {Steering, entanglement, nonlocality, and the {E}instein-{P}odolsky-{R}osen paradox},\ }\href {https://doi.org/10.1103/PhysRevLett.98.140402} {\bibfield  {journal} {\bibinfo  {journal} {Phys. Rev. Lett.}\ }\textbf {\bibinfo {volume} {98}},\ \bibinfo {pages} {140402} (\bibinfo {year} {2007})}\BibitemShut {NoStop}%
\bibitem [{\citenamefont {Bennett}\ \emph {et~al.}(1996)\citenamefont {Bennett}, \citenamefont {DiVincenzo}, \citenamefont {Smolin},\ and\ \citenamefont {Wootters}}]{bennett1996mixed}%
  \BibitemOpen
  \bibfield  {author} {\bibinfo {author} {\bibfnamefont {C.~H.}\ \bibnamefont {Bennett}}, \bibinfo {author} {\bibfnamefont {D.~P.}\ \bibnamefont {DiVincenzo}}, \bibinfo {author} {\bibfnamefont {J.~A.}\ \bibnamefont {Smolin}},\ and\ \bibinfo {author} {\bibfnamefont {W.~K.}\ \bibnamefont {Wootters}},\ }\bibfield  {title} {\bibinfo {title} {Mixed-state entanglement and quantum error correction},\ }\href {https://doi.org/10.1103/PhysRevA.54.3824} {\bibfield  {journal} {\bibinfo  {journal} {Phys. Rev. A}\ }\textbf {\bibinfo {volume} {54}},\ \bibinfo {pages} {3824} (\bibinfo {year} {1996})}\BibitemShut {NoStop}%
\bibitem [{\citenamefont {DiVincenzo}\ \emph {et~al.}(1999)\citenamefont {DiVincenzo}, \citenamefont {Fuchs}, \citenamefont {Mabuchi}, \citenamefont {Smolin}, \citenamefont {Thapliyal},\ and\ \citenamefont {Uhlmann}}]{divincenzo1999assistance}%
  \BibitemOpen
  \bibfield  {author} {\bibinfo {author} {\bibfnamefont {D.~P.}\ \bibnamefont {DiVincenzo}}, \bibinfo {author} {\bibfnamefont {C.~A.}\ \bibnamefont {Fuchs}}, \bibinfo {author} {\bibfnamefont {H.}~\bibnamefont {Mabuchi}}, \bibinfo {author} {\bibfnamefont {J.~A.}\ \bibnamefont {Smolin}}, \bibinfo {author} {\bibfnamefont {A.}~\bibnamefont {Thapliyal}},\ and\ \bibinfo {author} {\bibfnamefont {A.}~\bibnamefont {Uhlmann}},\ }\bibfield  {title} {\bibinfo {title} {Entanglement of assistance},\ }\href {https://doi.org/10.1007/3-540-49208-9_20} {\bibfield  {journal} {\bibinfo  {journal} {Lecture Notes in Computer Science}\ }\textbf {\bibinfo {volume} {1509}},\ \bibinfo {pages} {247} (\bibinfo {year} {1999})}\BibitemShut {NoStop}%
\bibitem [{\citenamefont {Vidal}(2000)}]{vidal2000monotones}%
  \BibitemOpen
  \bibfield  {author} {\bibinfo {author} {\bibfnamefont {G.}~\bibnamefont {Vidal}},\ }\bibfield  {title} {\bibinfo {title} {Entanglement monotones},\ }\href {https://doi.org/10.1080/09500340008244048} {\bibfield  {journal} {\bibinfo  {journal} {Journal of Modern Optics}\ }\textbf {\bibinfo {volume} {47}},\ \bibinfo {pages} {355} (\bibinfo {year} {2000})}\BibitemShut {NoStop}%
\bibitem [{\citenamefont {Horodecki}\ \emph {et~al.}(2003)\citenamefont {Horodecki}, \citenamefont {Horodecki}, \citenamefont {Horodecki}, \citenamefont {Horodecki}, \citenamefont {Oppenheim}, \citenamefont {Sen(De)},\ and\ \citenamefont {Sen}}]{horodecki2003reversible}%
  \BibitemOpen
  \bibfield  {author} {\bibinfo {author} {\bibfnamefont {M.}~\bibnamefont {Horodecki}}, \bibinfo {author} {\bibfnamefont {K.}~\bibnamefont {Horodecki}}, \bibinfo {author} {\bibfnamefont {P.}~\bibnamefont {Horodecki}}, \bibinfo {author} {\bibfnamefont {R.}~\bibnamefont {Horodecki}}, \bibinfo {author} {\bibfnamefont {J.}~\bibnamefont {Oppenheim}}, \bibinfo {author} {\bibfnamefont {A.}~\bibnamefont {Sen(De)}},\ and\ \bibinfo {author} {\bibfnamefont {U.}~\bibnamefont {Sen}},\ }\bibfield  {title} {\bibinfo {title} {Local information as a resource in distributed quantum systems},\ }\href {https://doi.org/10.1103/PhysRevLett.90.100402} {\bibfield  {journal} {\bibinfo  {journal} {Phys. Rev. Lett.}\ }\textbf {\bibinfo {volume} {90}},\ \bibinfo {pages} {100402} (\bibinfo {year} {2003})}\BibitemShut {NoStop}%
\bibitem [{\citenamefont {Brand\~ao}\ \emph {et~al.}(2013)\citenamefont {Brand\~ao}, \citenamefont {Horodecki}, \citenamefont {Oppenheim}, \citenamefont {Renes},\ and\ \citenamefont {Spekkens}}]{brandao2013resource}%
  \BibitemOpen
  \bibfield  {author} {\bibinfo {author} {\bibfnamefont {F.~G. S.~L.}\ \bibnamefont {Brand\~ao}}, \bibinfo {author} {\bibfnamefont {M.}~\bibnamefont {Horodecki}}, \bibinfo {author} {\bibfnamefont {J.}~\bibnamefont {Oppenheim}}, \bibinfo {author} {\bibfnamefont {J.~M.}\ \bibnamefont {Renes}},\ and\ \bibinfo {author} {\bibfnamefont {R.~W.}\ \bibnamefont {Spekkens}},\ }\bibfield  {title} {\bibinfo {title} {Resource theory of quantum states out of thermal equilibrium},\ }\href {https://doi.org/10.1103/PhysRevLett.111.250404} {\bibfield  {journal} {\bibinfo  {journal} {Phys. Rev. Lett.}\ }\textbf {\bibinfo {volume} {111}},\ \bibinfo {pages} {250404} (\bibinfo {year} {2013})}\BibitemShut {NoStop}%
\bibitem [{\citenamefont {Horodecki}\ and\ \citenamefont {Oppenheim}(2013)}]{horodecki2013fundamental}%
  \BibitemOpen
  \bibfield  {author} {\bibinfo {author} {\bibfnamefont {M.}~\bibnamefont {Horodecki}}\ and\ \bibinfo {author} {\bibfnamefont {J.}~\bibnamefont {Oppenheim}},\ }\bibfield  {title} {\bibinfo {title} {Fundamental limitations for quantum and nanoscale thermodynamics},\ }\href {https://doi.org/10.1038/ncomms3059} {\bibfield  {journal} {\bibinfo  {journal} {Nat. Commun.}\ }\textbf {\bibinfo {volume} {4}},\ \bibinfo {pages} {2059} (\bibinfo {year} {2013})}\BibitemShut {NoStop}%
\bibitem [{\citenamefont {Allahverdyan}\ \emph {et~al.}(2004)\citenamefont {Allahverdyan}, \citenamefont {Balian},\ and\ \citenamefont {Nieuwenhuizen}}]{allahverdyan2004maximal}%
  \BibitemOpen
  \bibfield  {author} {\bibinfo {author} {\bibfnamefont {A.~E.}\ \bibnamefont {Allahverdyan}}, \bibinfo {author} {\bibfnamefont {R.}~\bibnamefont {Balian}},\ and\ \bibinfo {author} {\bibfnamefont {T.~M.}\ \bibnamefont {Nieuwenhuizen}},\ }\bibfield  {title} {\bibinfo {title} {Maximal work extraction from finite quantum systems},\ }\href {https://doi.org/10.1209/epl/i2004-10101-2} {\bibfield  {journal} {\bibinfo  {journal} {EPL (Europhysics Letters)}\ }\textbf {\bibinfo {volume} {67}},\ \bibinfo {pages} {565} (\bibinfo {year} {2004})}\BibitemShut {NoStop}%
\bibitem [{\citenamefont {Alimuddin}\ \emph {et~al.}(2020)\citenamefont {Alimuddin}, \citenamefont {Guha},\ and\ \citenamefont {Parashar}}]{Alimuddin2020}%
  \BibitemOpen
  \bibfield  {author} {\bibinfo {author} {\bibfnamefont {M.}~\bibnamefont {Alimuddin}}, \bibinfo {author} {\bibfnamefont {T.}~\bibnamefont {Guha}},\ and\ \bibinfo {author} {\bibfnamefont {P.}~\bibnamefont {Parashar}},\ }\bibfield  {title} {\bibinfo {title} {Independence of work and entropy for equal-energetic finite quantum systems: Passive-state energy as an entanglement quantifier},\ }\href {https://doi.org/10.1103/PhysRevE.102.012145} {\bibfield  {journal} {\bibinfo  {journal} {Phys. Rev. E}\ }\textbf {\bibinfo {volume} {102}},\ \bibinfo {pages} {012145} (\bibinfo {year} {2020})}\BibitemShut {NoStop}%
\bibitem [{\citenamefont {Yang}\ \emph {et~al.}(2024)\citenamefont {Yang}, \citenamefont {Alimuddin}, \citenamefont {Yang},\ and\ \citenamefont {Luo}}]{yang2024ergotropicgap}%
  \BibitemOpen
  \bibfield  {author} {\bibinfo {author} {\bibfnamefont {X.}~\bibnamefont {Yang}}, \bibinfo {author} {\bibfnamefont {M.}~\bibnamefont {Alimuddin}}, \bibinfo {author} {\bibfnamefont {Y.-H.}\ \bibnamefont {Yang}},\ and\ \bibinfo {author} {\bibfnamefont {M.-X.}\ \bibnamefont {Luo}},\ }\bibfield  {title} {\bibinfo {title} {Characterizing bipartite entanglement via the ergotropic gap},\ }\href {https://doi.org/10.1007/s11128-024-04599-0} {\bibfield  {journal} {\bibinfo  {journal} {Quantum Information Processing}\ }\textbf {\bibinfo {volume} {23}},\ \bibinfo {pages} {382} (\bibinfo {year} {2024})}\BibitemShut {NoStop}%
\bibitem [{\citenamefont {Baumgratz}\ \emph {et~al.}(2014)\citenamefont {Baumgratz}, \citenamefont {Cramer},\ and\ \citenamefont {Plenio}}]{baumgratz2014quantifying}%
  \BibitemOpen
  \bibfield  {author} {\bibinfo {author} {\bibfnamefont {T.}~\bibnamefont {Baumgratz}}, \bibinfo {author} {\bibfnamefont {M.}~\bibnamefont {Cramer}},\ and\ \bibinfo {author} {\bibfnamefont {M.~B.}\ \bibnamefont {Plenio}},\ }\bibfield  {title} {\bibinfo {title} {Quantifying coherence},\ }\href {https://doi.org/10.1103/PhysRevLett.113.140401} {\bibfield  {journal} {\bibinfo  {journal} {Phys. Rev. Lett.}\ }\textbf {\bibinfo {volume} {113}},\ \bibinfo {pages} {140401} (\bibinfo {year} {2014})}\BibitemShut {NoStop}%
\bibitem [{\citenamefont {Winter}\ and\ \citenamefont {Yang}(2016)}]{winter2016operational}%
  \BibitemOpen
  \bibfield  {author} {\bibinfo {author} {\bibfnamefont {A.}~\bibnamefont {Winter}}\ and\ \bibinfo {author} {\bibfnamefont {D.}~\bibnamefont {Yang}},\ }\bibfield  {title} {\bibinfo {title} {Operational resource theory of coherence},\ }\href {https://doi.org/10.1103/PhysRevLett.116.120404} {\bibfield  {journal} {\bibinfo  {journal} {Phys. Rev. Lett.}\ }\textbf {\bibinfo {volume} {116}},\ \bibinfo {pages} {120404} (\bibinfo {year} {2016})}\BibitemShut {NoStop}%
\bibitem [{\citenamefont {Blackwell}(1953)}]{blackwell1953equivalent}%
  \BibitemOpen
  \bibfield  {author} {\bibinfo {author} {\bibfnamefont {D.}~\bibnamefont {Blackwell}},\ }\bibfield  {title} {\bibinfo {title} {Equivalent comparisons of experiments},\ }\href {https://doi.org/10.1214/aoms/1177729032} {\bibfield  {journal} {\bibinfo  {journal} {Ann. Math. Stat.}\ }\textbf {\bibinfo {volume} {24}},\ \bibinfo {pages} {265} (\bibinfo {year} {1953})}\BibitemShut {NoStop}%
\bibitem [{\citenamefont {Shmaya}(2005)}]{shmaya2005comparison}%
  \BibitemOpen
  \bibfield  {author} {\bibinfo {author} {\bibfnamefont {E.}~\bibnamefont {Shmaya}},\ }\bibfield  {title} {\bibinfo {title} {Comparison of information structures and completely positive maps},\ }\href {https://doi.org/10.1088/0305-4470/38/44/013} {\bibfield  {journal} {\bibinfo  {journal} {Journal of Physics A: Mathematical and General}\ }\textbf {\bibinfo {volume} {38}},\ \bibinfo {pages} {9717} (\bibinfo {year} {2005})}\BibitemShut {NoStop}%
\bibitem [{\citenamefont {Wootters}(1998)}]{wootters1998}%
  \BibitemOpen
  \bibfield  {author} {\bibinfo {author} {\bibfnamefont {W.~K.}\ \bibnamefont {Wootters}},\ }\bibfield  {title} {\bibinfo {title} {Entanglement of formation of an arbitrary state of two qubits},\ }\href {https://doi.org/10.1103/PhysRevLett.80.2245} {\bibfield  {journal} {\bibinfo  {journal} {Phys. Rev. Lett.}\ }\textbf {\bibinfo {volume} {80}},\ \bibinfo {pages} {2245} (\bibinfo {year} {1998})}\BibitemShut {NoStop}%
\bibitem [{\citenamefont {Matsumoto}(2014)}]{matsumoto2014example}%
  \BibitemOpen
  \bibfield  {author} {\bibinfo {author} {\bibfnamefont {K.}~\bibnamefont {Matsumoto}},\ }\bibfield  {title} {\bibinfo {title} {An example of a quantum statistical model which cannot be mapped to a less informative one by any trace preserving positive map},\ }\href {https://doi.org/10.48550/arXiv.1409.5658} {\bibfield  {journal} {\bibinfo  {journal} {arXiv:1409.5658}\ } (\bibinfo {year} {2014})}\BibitemShut {NoStop}%
\bibitem [{\citenamefont {Gurvits}(2003)}]{gurvits2003}%
  \BibitemOpen
  \bibfield  {author} {\bibinfo {author} {\bibfnamefont {L.}~\bibnamefont {Gurvits}},\ }\href {https://arxiv.org/abs/quant-ph/0303055} {\bibinfo {title} {Classical deterministic complexity of edmonds' problem and quantum entanglement}} (\bibinfo {year} {2003}),\ \Eprint {https://arxiv.org/abs/quant-ph/0303055} {arXiv:quant-ph/0303055 [quant-ph]} \BibitemShut {NoStop}%
\bibitem [{\citenamefont {Gharibian}(2009)}]{gharibian2009}%
  \BibitemOpen
  \bibfield  {author} {\bibinfo {author} {\bibfnamefont {S.}~\bibnamefont {Gharibian}},\ }\href {https://arxiv.org/abs/0810.4507} {\bibinfo {title} {Strong np-hardness of the quantum separability problem}} (\bibinfo {year} {2009}),\ \Eprint {https://arxiv.org/abs/0810.4507} {arXiv:0810.4507 [quant-ph]} \BibitemShut {NoStop}%
\bibitem [{\citenamefont {Wolf}\ and\ \citenamefont {P\'erez-Garc\'ia}(2007)}]{wolf2007quantum}%
  \BibitemOpen
  \bibfield  {author} {\bibinfo {author} {\bibfnamefont {M.~M.}\ \bibnamefont {Wolf}}\ and\ \bibinfo {author} {\bibfnamefont {D.}~\bibnamefont {P\'erez-Garc\'ia}},\ }\bibfield  {title} {\bibinfo {title} {Quantum capacities of channels with small environment},\ }\href {https://doi.org/10.1103/PhysRevA.75.012303} {\bibfield  {journal} {\bibinfo  {journal} {Physical Review A}\ }\textbf {\bibinfo {volume} {75}},\ \bibinfo {pages} {012303} (\bibinfo {year} {2007})}\BibitemShut {NoStop}%
\bibitem [{\citenamefont {Cubitt}\ \emph {et~al.}(2008)\citenamefont {Cubitt}, \citenamefont {Ruskai},\ and\ \citenamefont {Smith}}]{cubitt2008}%
  \BibitemOpen
  \bibfield  {author} {\bibinfo {author} {\bibfnamefont {T.~S.}\ \bibnamefont {Cubitt}}, \bibinfo {author} {\bibfnamefont {M.~B.}\ \bibnamefont {Ruskai}},\ and\ \bibinfo {author} {\bibfnamefont {G.}~\bibnamefont {Smith}},\ }\bibfield  {title} {\bibinfo {title} {The structure of degradable quantum channels},\ }\href {https://doi.org/10.1063/1.2953685} {\bibfield  {journal} {\bibinfo  {journal} {J. Math. Phys.}\ }\textbf {\bibinfo {volume} {49}},\ \bibinfo {pages} {102104} (\bibinfo {year} {2008})},\ \Eprint {https://arxiv.org/abs/0802.1360} {arXiv:0802.1360} \BibitemShut {NoStop}%
\bibitem [{\citenamefont {Shirokov}(2010)}]{shirokov2010properties}%
  \BibitemOpen
  \bibfield  {author} {\bibinfo {author} {\bibfnamefont {M.~E.}\ \bibnamefont {Shirokov}},\ }\bibfield  {title} {\bibinfo {title} {On properties of the space of quantum states and their application to the construction of entanglement monotones},\ }\href@noop {} {\bibfield  {journal} {\bibinfo  {journal} {Izvestiya: Mathematics}\ }\textbf {\bibinfo {volume} {74}},\ \bibinfo {pages} {849} (\bibinfo {year} {2010})}\BibitemShut {NoStop}%
\bibitem [{\citenamefont {Koashi}\ and\ \citenamefont {Winter}(2004)}]{koashi2004monogamy}%
  \BibitemOpen
  \bibfield  {author} {\bibinfo {author} {\bibfnamefont {M.}~\bibnamefont {Koashi}}\ and\ \bibinfo {author} {\bibfnamefont {A.}~\bibnamefont {Winter}},\ }\bibfield  {title} {\bibinfo {title} {Monogamy of quantum entanglement and other correlations},\ }\href {https://doi.org/10.1103/PhysRevA.69.022309} {\bibfield  {journal} {\bibinfo  {journal} {Phys. Rev. A}\ }\textbf {\bibinfo {volume} {69}},\ \bibinfo {pages} {022309} (\bibinfo {year} {2004})}\BibitemShut {NoStop}%
\bibitem [{\citenamefont {Luo}(2008)}]{luo2008quantum}%
  \BibitemOpen
  \bibfield  {author} {\bibinfo {author} {\bibfnamefont {S.}~\bibnamefont {Luo}},\ }\bibfield  {title} {\bibinfo {title} {Quantum discord for two-qubit systems},\ }\href {https://doi.org/10.1103/PhysRevA.77.042303} {\bibfield  {journal} {\bibinfo  {journal} {Physical Review A}\ }\textbf {\bibinfo {volume} {77}},\ \bibinfo {pages} {042303} (\bibinfo {year} {2008})}\BibitemShut {NoStop}%
\bibitem [{\citenamefont {Paddock}\ and\ \citenamefont {Chen}(2017)}]{paddock2017characterization}%
  \BibitemOpen
  \bibfield  {author} {\bibinfo {author} {\bibfnamefont {C.}~\bibnamefont {Paddock}}\ and\ \bibinfo {author} {\bibfnamefont {J.}~\bibnamefont {Chen}},\ }\bibfield  {title} {\bibinfo {title} {A characterization of antidegradable qubit channels},\ }\href {https://doi.org/10.48550/arXiv.1712.03399} {\bibfield  {journal} {\bibinfo  {journal} {arXiv:1712.03399}\ } (\bibinfo {year} {2017})}\BibitemShut {NoStop}%
\bibitem [{\citenamefont {Zhu}\ and\ \citenamefont {Wang}(2026)}]{zhu2026zero}%
  \BibitemOpen
  \bibfield  {author} {\bibinfo {author} {\bibfnamefont {C.}~\bibnamefont {Zhu}}\ and\ \bibinfo {author} {\bibfnamefont {X.}~\bibnamefont {Wang}},\ }\href {https://arxiv.org/abs/2609.10520} {\bibinfo {title} {Private communication via zero-private-capacity quantum channels}} (\bibinfo {year} {2026}),\ \Eprint {https://arxiv.org/abs/2609.10520} {arXiv:2609.10520 [quant-ph]} \BibitemShut {NoStop}%
\end{thebibliography}
\end{document}